\documentclass[11pt]{article}    
\usepackage{graphicx}
\usepackage{algorithm}
\usepackage{algpseudocode}
\usepackage{url}
\usepackage{authblk}
\usepackage[margin=1in]{geometry}
\usepackage[hyperpageref]{backref}
\usepackage[utf8]{inputenc}
\usepackage{amsfonts}
\usepackage{amsmath}
\usepackage{dblfloatfix}
\usepackage{bm}
\usepackage{mdframed}
\usepackage{booktabs}
\usepackage{tabularx}
\usepackage[dvipsnames]{xcolor} 
\usepackage{natbib}
\usepackage[driverfallback=dvipdfm, colorlinks=true]{hyperref}
\usepackage{epstopdf}
\usepackage{bbm}

\newtheorem{proposition}{Proposition}[section]
\newtheorem{corollary}{Corollary}[section]

\begin{document}\sloppy

\title{Anytime Parallel Tempering}



\author[$1$]{Alix Marie d'Avigneau\thanks{A. Marie d'Avigneau is supported by the UK Engineering and Physical Sciences Research Council (EPSRC).}}
\author[$1$]{Sumeetpal S. Singh}
\author[$2$]{Lawrence M. Murray}
\affil[$1$]{\small Signal Processing and Communications Group, Department of Engineering, University of Cambridge, Cambridge, UK}
\affil[$2$]{\small Uber AI Labs, San Francisco, CA, USA}
\date{}
\maketitle

\begin{abstract}
	Developing efficient MCMC algorithms is indispensable in Bayesian inference. In parallel tempering, multiple interacting MCMC chains run to more efficiently explore the state space and improve performance. The multiple chains advance independently through local moves, and the performance enhancement steps are exchange moves, where the chains pause to exchange their current sample amongst each other. To accelerate the independent local moves, they may be performed simultaneously on multiple processors. Another problem is then encountered: depending on the MCMC implementation and inference problem, local moves can take a varying and random amount of time to complete. There may also be infrastructure-induced variations, such as competing jobs on the same processors, which arises in cloud computing. 
	Before exchanges can occur, all chains must complete the local moves they are engaged in to avoid introducing a potentially substantial bias  (Proposition \ref{prop:anytimefull}).  
	To solve this issue of randomly varying local move completion times in multi-processor parallel tempering, we adopt the Anytime Monte Carlo framework of \cite{murray2016anytime}: we impose real-time deadlines on the parallel local moves and perform exchanges at these deadlines without any processor idling. We show our methodology for exchanges at real-time deadlines does not introduce a bias and leads to significant performance enhancements over the na\"{i}ve approach of idling until every processor's local moves complete. The methodology is then applied in an ABC setting, where an Anytime ABC parallel tempering algorithm is derived for the difficult task of estimating the parameters of a Lotka-Volterra predator-prey model, and similar efficiency enhancements are observed. 
\end{abstract}

\section{Introduction}
\label{intro}
Consider a set of $m$ observations $y = \left\{y_1, \ldots, y_m\right\} \in \mathcal{Y}$ following a probability model with underlying parameters $\theta \in \Theta$ and associated \textit{likelihood} $f(y_1, \ldots, y_m\,|\,\theta)$ which we abbreviate to $f(y \,|\, \theta)$. In most cases, the posterior $\pi(\text{d}\theta)$ of interest is intractable and must be approximated using computational tools such as the commonly used \textit{Metropolis}-\textit{Hastings} (M-H) algorithm (\cite{robertg}) with random walk proposals, for example. However, as models become more complex, the exploration of the posterior using such basic methods quickly becomes inefficient (\cite{beskos2009optimal}). Furthermore, the model itself can pose its own challenges such as the likelihood becoming increasingly costly or even impossible to evaluate (\cite{tavare1997inferring}); the Lotka-Volterra predator-prey model of Section \ref{cha:exp} is a concrete example. \par
\textit{Parallel tempering}, initially proposed by \cite{swendsen1986replica} and further developed under the name Metropolis-coupled Markov chain Monte Carlo (MC)$^3$ by \cite{geyer1991markov}, is a generic method for improving the efficiency of MCMC that can be very effective without significantly altering the original MCMC algorithm, beyond perhaps tuning its local proposals for each temperature. The parallel tempering algorithm runs multiple interacting MCMC chains to more efficiently explore the state space. The multiple MCMC chains are advanced independently, in what is known as the local moves, and the performance enhancement steps are the exchange moves, where the chains pause and attempt to swap their current sample amongst each other.  
Parallel tempering allows for steps of various sizes to be made when exploring the parameter space, which makes the algorithm effective, even when the distribution we wish to sample from has multiple modes. In order to reduce the real time taken to perform the independent local moves, they may be performed simultaneously on multiple processors, a feature we will focus on in this work. \par
Let the parallel tempering MCMC chain be $\left(X_n^{1:\Lambda}\right)_{n=1}^{\infty}  = \left(X_n^1, \ldots, X_n^{\Lambda}\right)_{n=1}^{\infty}$ with initial state $\left(X^{1:\Lambda}_0\right)$ and target distribution	
\begin{equation}
	\label{eq:pt}
	\pi(\text{d}x^{1:\Lambda}) \propto \prod_{\lambda=1}^{\Lambda} \pi_{\lambda}(\text{d}x^{\lambda}),
\end{equation}	
where the $\pi_{\lambda}(\, \cdot \,)$ are independent marginals corresponding to the target distribution of each of $\Lambda$ chains, running in parallel at different temperatures indexed by $\lambda$. One of these chains, say $\lambda=\Lambda$, is the \textit{cold} chain, and its target distribution $\pi_{\Lambda} = \pi$ is the posterior of interest. At each step $n$ of parallel tempering (\cite{geyer2011importance}), one of two types of updates is used to advance the Markov chain $X_n^{1:\Lambda}$ to its next state:
\begin{enumerate}
	\item Independent \textit{local moves}: for example, a standard Gibbs or Metropolis-Hastings update, applied to each tempered chain $X^{\lambda}_n$ in parallel. 
	\item Interacting \textit{exchange moves}: propose to swap the states $x \sim \pi_{\lambda}$ and $x' \sim \pi_{\lambda'}$ of one or more pairs of adjacent chains. For each pair, accept a swap with probability
	\begin{equation}
		\label{eq:swap}
		a_{swap}(x', x) = \min\left\{1, \frac{\pi_{\lambda}(x') \pi_{\lambda'}(x)}{\pi_{\lambda}(x) \pi_{\lambda'}(x')}\right\}, 
	\end{equation}
	otherwise, the chains in the pair retain their current states.
\end{enumerate}	
With the cold chain providing the desired precision and the warmer chains more freedom of movement when exploring the parameter space, the combination of the two types of update allows all chains to mix much faster than any one of them would mix on its own. This provides a way to jump from mode to mode in far fewer steps than would be required under a standard non-tempered implementation using, say, the Metropolis-Hastings algorithm.\par

A particular advantage of parallel tempering is that it is possible to perform the independent local moves in parallel on multiple processors in order to reduce the real time taken to complete them. Unfortunately, this gives rise to the following  problem:  depending on the MCMC implementation and the inference problem itself, the local moves can take a \emph{varying and random} amount of time to complete, which depends on the part of the state space it is exploring  (see the Lotka-Volterra predator-prey model in Section \ref{sec:ABC:LV} for a specific real example). Thus, before the exchange moves can occur, all chains \emph{must} complete the local move they are engaged in to avoid introducing a potentially substantial bias (see Proposition \ref{prop:anytimefull}). Additionally, the time taken to complete local moves may also reflect computing infrastructure induced variations, for example, due to variations in processor hardware, memory bandwidth, network traffic, I/O load, competing jobs on the same processors, as well as potential unforeseen interruptions, all of which affect the compute time of local moves. Local moves in parallel tempering algorithms can also have temperature-dependent completion times. This is the case of the approximate Bayesian computation (ABC) application in Section \ref{cha:APTMC:ABC}. In \cite{earl2004optimal}, the authors consider a similar problem of temperature $\lambda$ dependent real completion times of local moves. To tackle the problem, they redistribute the chains among the processors in order to minimise processor idling that occurs while waiting for all local moves to finish. This strategy is a deterministic allocation of processor time to simulation and entails completing part of a simulation on one processor and then continuing on another. Our approach to removing idling doesn't involve redistributing partially completed simulations, and instead imposes real-time deadlines at which simulations are stopped to perform exchange moves before resuming work on their respective processors. The contributions of this paper are as follows. \par

Firstly, to solve the problem of randomly distributed local move completion times when parallel tempering is implemented on a multi-processor computing resource, we adopt the Anytime Monte Carlo framework of \cite{murray2016anytime}: we guarantee the simultaneous readiness of all chains by imposing real-time deadlines on the parallelly computed local moves, and perform exchange moves at these deadlines without any idling, i.e. without waiting for the slowest of them to complete their local moves. Idling is both a financial cost, for example in a cloud computing setting, and can also significantly reduce the effective Monte Carlo sample size returned. We show that hard deadlines introduce a bias which we mitigate using the Anytime framework (see Proposition \ref{prop:anytime:exch}). \par

Secondly, we illustrate our gains through detailed numerical work. The first experiment considered is a mixture model where the biased and de-biased target distributions can be characterised for ease of comparison with the numerical results.
We then apply our Anytime parallel tempering  methodology in the realm of ABC (\cite{tavare1997inferring, pritchard1999population}). In ABC, simulation is used instead of likelihood evaluations, which makes it particularly useful for Bayesian problems where the likelihood is unavailable or too costly to compute. In \cite{lee2012choice}, a more efficient MCMC kernel for ABC (as measured by the effective sample size), called the \textit{1-hit MCMC kernel}, was devised to significantly improve the probability that a good proposal in the direction of a higher posterior density is accepted, thus more closely mimicking exact likelihood evaluations. This new MCMC kernel was subsequently shown in \cite{lee2014variance} to also theoretically outperform competing ABC methods. The 1-hit kernel has a random execution time that depends on the part of the parameter space being explored, and is thus a good candidate for our Anytime parallel tempering method. 	
In this paper, we show that we can improve the performance of the 1-hit MCMC kernel by introducing tempering and exchange moves, and embed the resulting parallel tempering algorithm within the Anytime framework to mitigate processor idling due to random local move completion times. Parallel tempering for ABC has been proposed by \cite{baragatti2013likelihood}, but hasn't been studied in the Anytime context as we do for random local move completion times, nor has the more efficient 1-hit MCMC kernel been employed. We perform a detailed numerical study of the Lotka-Volterra predator-prey model, which has an intractable likelihood and is a popular example used to contrast methods in the ABC literature  (\cite{fearnhead2012constructing, toni2009approximate, prangle2017adapting}). The time taken to simulate from the Lotka-Volterra model is random and parameter value dependent; this randomness is in addition to that induced by the 1-hit kernel. 	 \par		 
The Anytime parallel tempering framework can be applied in several contexts. For example, another candidate for our framework is reversible jump MCMC (RJ-MCMC) by \cite{green1995reversible}, which is a variable-dimension Bayesian model inference algorithm.  An instance of RJ-MCMC within a parallel tempering algorithm is given in \cite{jasra2007population}, where multiple chains are simultaneously updating states of variable dimensions (depending on the model currently considered on each chain), and the real completion time of local moves depends on the dimension of the state space under the current model. Additionally, in the fixed dimension parallel tempering setting, if the local moves use any of the following MCMC kernels, then they have a parameter dependent completion time and thus could benefit from an Anytime formulation: the no-U-turn sampler (NUTS) (\cite{hoffman2014no}) and elliptical slice sampling (\cite{murray2010elliptical, nishihara2014parallel}). Even if the local moves do not take a variable random time to complete by design (\cite{friel2008marginal, calderhead2009estimating}), computer infrastructure induced variations, such as memory bandwidth, competing jobs, etc. can still affect the real completion time of local moves in a parallel tempering algorithm, such as in \cite{rodinger2006distributed}. In the statistical mechanics literature, there are also parallel tempering-based simulation problems where the local move completion time is temperature- and parameter-dependent as well as random, e.g. see \cite{hritz2007optimization, karimi2011high, wang2003parallel, earl2004optimal}, and thus could benefit from our Anytime formulation. Finally, the Anytime framework has not been tested beyond the SMC$^2$ example of \cite{murray2016anytime}, but it can be applied to any parallelisable population-based MCMC algorithm which includes local moves and interacting moves where all processors must communicate, such as  sequential Monte Carlo (SMC) samplers (\cite{del2006sequential}), or parallelised generalised elliptical slice sampling (\cite{nishihara2014parallel}).\par
This paper is structured as follows. Sections  \ref{cha:literature} and \ref{cha:APTMC} develop our Anytime Parallel Tempering Monte Carlo (APTMC) algorithm and then Section \ref{cha:APTMC:ABC} extends our framework further for the 1-hit MCMC kernel of \cite{lee2012choice} for ABC. Experiments are run in Section \ref{cha:exp} and include a carefully constructed synthetic example to demonstrate the workings and salient features of Anytime parallel tempering.  Section \ref{cha:exp} also presents an application of Anytime parallel tempering to the problem of estimating the parameters of a stochastic Lotka-Volterra predator-prey model. Finally, Section \ref{cha:discussion} provides a summary and some concluding remarks.

\section{Anytime Monte Carlo}
\label{cha:literature}
Let $\left(X_n\right)_{n=0}^{\infty}$ be a Markov chain with initial state $X_0$, evolving on state space $\mathcal{X}$, with transition kernel $X_n \, | \, x_{n-1} \sim \kappa(\text{d}x_n | x_{n-1})$ and target distribution $\pi(\text{d}x)$. Define the \textit{hold time} $H_{n-1}$ as the random and positive real time required to complete the computations necessary to transition from state $X_{n-1}$ to $X_n$ via the kernel $\kappa$. Then let $H_{n-1} \, | \, x_{n-1} \sim \tau(\text{d}h_{n-1}|x_{n-1})$ where $\tau$ is the hold time distribution.\par
\begin{figure*}[htbp]
	\centering
	\includegraphics[width=\textwidth]{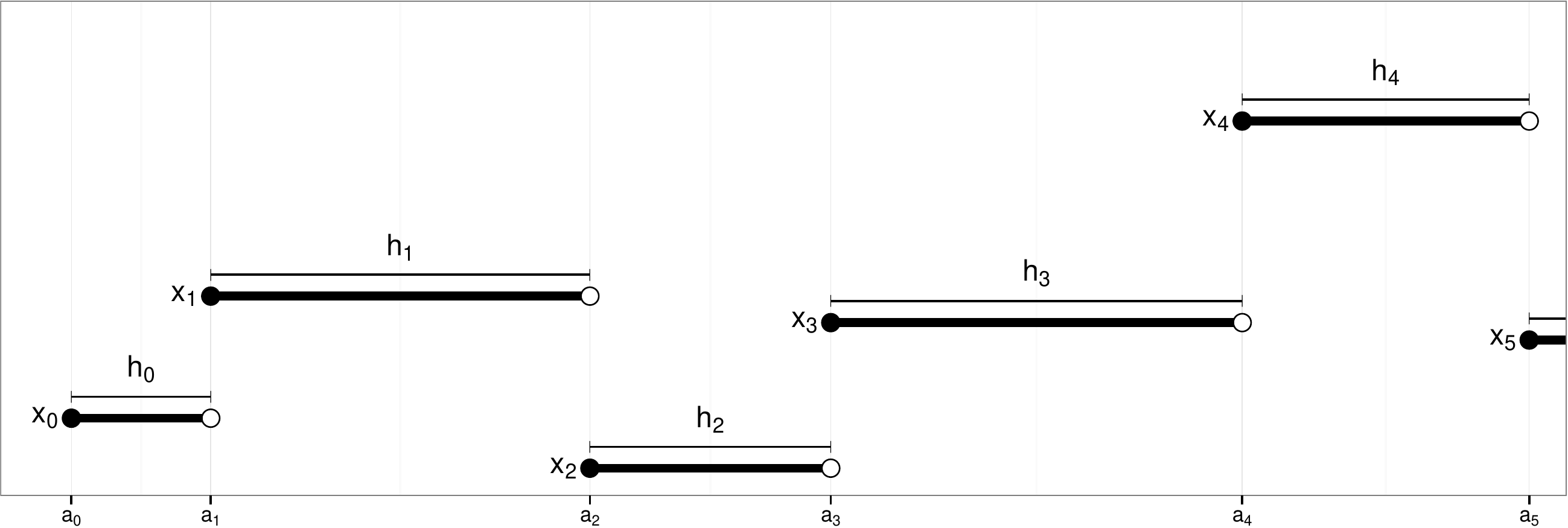}
	\caption{(\cite{murray2016anytime}, Figure 1) Real-time realisation of a Markov chain with states $\left(X_n\right)_{n=0}^{\infty}$, arrival times $\left(A_n\right)_{n=0}^{\infty}$ and hold times $\left(H_n\right)_{n=0}^{\infty}$ .}
	\label{fig:AMC}
\end{figure*}
Assume that the hold time $H > \epsilon >0$ for minimal time $\epsilon$, $\sup_{x \in \mathcal{X}} \mathbb{E}\left[H\,|\,x\right] < \infty$, and the hold time distribution $\tau$ is homogeneous in time. In general, nothing is known about the hold time distribution $\tau$ except how to sample from it, i.e. by recording the time taken by the algorithm to simulate $X_n \, | \, x_{n-1}$. Let $\kappa(\text{d}x_n, \text{d}h_{n-1} | x_{n-1}) = \kappa(\text{d}x_n | h_{n-1}, x_{n-1}) \tau(\text{d}h_{n-1}|x_{n-1})$ be a joint kernel. The transition kernel $\kappa(\text{d}x_n | x_{n-1})$ is the marginal of the joint kernel over all possible hold times $H_{n-1}$. Denote by $\left(X_n\right)_{n=0}^{\infty}$ and $\left(H_n\right)_{n=0}^{\infty}$ the states and hold times of the joint process, and define the \textit{arrival time} of the $n$-th state as
\begin{equation*}
	A_n := \sum_{i=0}^{n-1} H_i, \qquad n \geq 1,
\end{equation*} 
where $a_0 := 0$. A possible realisation of the joint process is illustrated in Figure \ref{fig:AMC}. \par 
Let the process $N(t) := \sup \left\{n:A_n\leq t \right\}$ count the number of arrivals by time $t$. From this, construct a continuous Markov jump process $\left(X, L\right)(t)$ where $X(t) := X_{N(t)}$ and $L(t):=t-A_{N(t)}$ is the \textit{lag time} elapsed since the last jump. This continuous process describes the progress of the computation in real time. 
\begin{proposition} 
	\label{prop:anytimefull}
	(\cite{murray2016anytime}, Proposition 1) The continuous Markov jump process $\left(X, L\right)(t)$ has stationary distribution given by
	\begin{equation}
		\label{eq:anytimefull}
		\alpha(\mathrm{d}x, \mathrm{d}l) = \frac{\bar{F}_{\tau}\left(l \, | \, x \right)}
		{\mathbb{E}\left[H\right]} \pi(\mathrm{d}x) \mathrm{d}l,
	\end{equation}
\end{proposition}
where $\bar{F}_{\tau}\left(l \, | \, x \right) = 1-F_{\tau}\left(l \, | \, x \right)$, and $F_{\tau}\left(l \, | \, x \right)$ is the cumulative distribution function (cdf) of $\tau(\text{d}h_n|x_{n})$.
\begin{corollary}
	\label{prop:anytime}
	(\cite{murray2016anytime}, Corollary 2) The marginal $\alpha(\mathrm{d}x)$ of the density in \eqref{eq:anytimefull} is length-biased with respect to the target density $\pi(\mathrm{d}x)$ by expected hold time, i.e.
	\begin{equation}
		\label{eq:anytime}
		\alpha(\mathrm{d}x) = \frac{\mathbb{E}\left[H  \, | \, x\right]}
		{\mathbb{E}\left[H\right]} \pi(\mathrm{d}x).
	\end{equation}
\end{corollary}
The proofs of Proposition \ref{prop:anytimefull} and Corollary \ref{prop:anytime} are given in \cite{murray2016anytime}.\par
The distribution $\alpha$ is referred to as the \textit{anytime distribution} and is the stationary distribution of the Markov jump process. Note that Proposition \ref{prop:anytime} suggests that when the real time taken to draw a sample depends on the state of the Markov chain, i.e. $\mathbb{E}[H\,|\,x]\neq \mathbb{E}[H]$, a length bias with respect to computation time is introduced.  In other words, when interrupted at real time $t$, the state of a Monte Carlo computation targeting $\pi$ is distributed according to the anytime distribution $\alpha$, which can essentially be seen as a length-biased target distribution. 
This bias diminishes with time, and when an empirical approximation or average over all post burn-in samples is required, it may be rendered negligible for a long enough computation. However, the bias in the final state does not diminish with time, and when this final state is important $-$ which is the case in parallel tempering $-$ the bias cannot be avoided by running the algorithm for longer. We now discuss the approach in \cite{murray2016anytime} to correct this bias. The main idea is to make it so expected hold time is independent of $X$, which leads to $\mathbb{E}\left[H\,|\,x\right] = \mathbb{E}\left[H\right]$ and hence $\alpha(\text{d}x) = \pi(\text{d}x)$, following Corollary \ref{prop:anytime}. This is trivially the case for iid sampling as $\kappa(\text{d}x|x_{n-1})=\pi(\text{d}x)$, so the hold time $H_{n-1}$ for $X_{n-1}$ is the time taken to sample $X_n \sim \pi(\text{d}x)$, and therefore independent of the state $X_{n-1}$. One approach to non-iid sampling involves simulating $K+1$ Markov chains for $K > 0$, where we assume for now that all the Markov chains are targeting $\pi$ and using the same transition kernel $\kappa$ and hold time distribution $\tau$. These $K+1$ chains are simulated on the same processor in a serial schedule. This ensures that whenever the real-time deadline $t$ is reached, states from all but one of the chains, say the $(K+1)$-th chain, are independently distributed according to the target $\pi$.  Since the $(K+1)$-th chain is the currently working chain, i.e. the latest to go through the simulation process, its state at the real-time deadline is distributed according to the anytime distribution $\alpha$. Simply discarding or ignoring the state of this $(K+1)$-th chain eliminates the length bias. See \cite{murray2016anytime} (Section 2.1) for more details.\par
Using this multiple chain construction, it is thus possible draw samples from $\pi$ by interrupting the process at any time $t$. This sets the basis for the focus of this paper: the Anytime Parallel Tempering Monte Carlo (APTMC) algorithm, described next. From this point onward, the number of chains on a given worker or processor within the Anytime framework is referred to as $K$ rather than $K+1$ for simplicity.

\section{Anytime Parallel Tempering Monte Carlo (APTMC)} \label{cha:APTMC}
		
	\subsection{Overview}
	Consider the problem in which we wish to sample from target distribution $\pi(\text{d}x)$. In a parallel tempering framework, construct $\Lambda$ Markov chains where each individual chain $\lambda$ targets the tempered distribution	
	\begin{equation*}
		\pi_{\lambda}(\text{d}x) \propto \pi(\text{d}x)^{\frac{\lambda}{\Lambda}}
	\end{equation*}	
	and is associated with kernel $\kappa_{\lambda}(\text{d}x_{n} \, | \, \text{d}x_{n-1})$ and hold time distribution $\tau_{\lambda}(\text{d}h_{n}\,|\,x_{n})$. In this setting, the hold time distribution is not assumed to be homogeneous across all chains, and may be temperature-dependent. Assume that all $\Lambda$ chains are running concurrently on $\Lambda$ processors. We aim to interrupt the computations on a real-time schedule of times $t_1, t_2, t_3, \ldots$ to perform exchange moves between adjacent pairs of chains before resuming the local moves. To illustrate the challenge of this task, we discuss the case where $\Lambda=2$. Let $\pi_2$ be the desired posterior and $\pi_1$ the `warm' chain, with associated hold time distributions $\tau_1$ and $\tau_2$, respectively. When the two chains are interrupted at some time $t$, assume that the current sample on chain $1$ is $X_m^{1}$ and that of chain $2$ is $X_n^2$. It follows from Corollary \ref{prop:anytime} that
	\begin{equation*}
		X_m^1 \sim \alpha_1(\text{d}x) = \frac{\mathbb{E}\left[H_1\,|\,x\right]}
		{\mathbb{E}\left[H_1\right]}\pi_1(\text{d}x) \neq \pi_1(\text{d}x),
	\end{equation*}	
	and similarly for $X_n^2$. Exchanging the samples using the acceptance probability in \eqref{eq:swap} is incorrect. Indeed, exchanging using the current samples $X_m^1$ and $X_n^2$, if accepted, will result in the sample sets $\left\{X_1^1, X_2^1, \ldots \right\}$ and $\left\{X_1^2, X_2^2, \ldots \right\}$ being corrupted with samples which arise from their respective length-biased, anytime distributions $\alpha_1$ and $\alpha_2$, as opposed to being exclusively from $\pi_1$ and $\pi_2$. Furthermore, the expressions for $\alpha_1$ and $\alpha_2$ will most often be unavailable, since their respective hold time distributions $\tau_1$ and $\tau_2$ are not explicitly known but merely implied by the algorithm used to simulate the two chains. Finally, we could wait for chains $1$ and $2$ complete their computation of $X_{m+1}^1$ and $X_{n+1}^2$ respectively, and then accept/reject the exchange $\left(X_{m+1}^1, X_{n+1}^2\right) \rightarrow \left(X_{n+1}^2, X_{m+1}^1\right)$ according to \eqref{eq:swap}. This approach won't introduce a bias but can result in one processor idling while the slower computation finishes. We show this can result in significant idling in numerical examples.\par
	In the next section, we describe how to correctly implement exchange moves within the Anytime framework.
	\subsection{Anytime exchange moves} \label{sec:APTMC:exch}
	Here, we adapt the multi-chain construction devised to remove the bias present when sampling from $\Lambda$ Markov chains, where each chain $\lambda$ targets the distribution $\pi_{\lambda}$  for $\lambda = 1, \ldots, \Lambda$. Associated with each chain is MCMC kernel $\kappa_{\lambda}(\text{d}x^{\lambda}_n \, | \, \text{d}x^{\lambda}_{n-1})$ and hold time distribution $\tau_{\lambda}(\text{d}h \, | \, x)$. 
	\begin{proposition}
		\label{prop:anytime:exch}
		Let $\pi_\lambda(\mathrm{d}x)$, $\lambda = 1\ldots, \Lambda$ be the stationary distributions of $\Lambda$ Markov chains with associated MCMC kernels $\kappa_{\lambda}(\mathrm{d}x^{\lambda}_n \, | \, \mathrm{d}x^{\lambda}_{n-1})$ and hold time distributions $\tau_{\lambda}(\mathrm{d}h \, | \, x)$. Assume the chains are updated sequentially and let $j$ be the index of the currently working chain. The joint anytime distribution is the following generalisation of Proposition \ref{prop:anytime}
		\begin{align*}
			&A(\mathrm{d}x^{1:\Lambda}, \mathrm{d}l, j) 
			= \frac{1}{\Lambda} \frac{\mathbb{E}\left[H \, | \, j\right]}{\mathbb{E}\left[H\right]} \alpha_j(\mathrm{d}x^{j}, \mathrm{d}l) \prod_{\lambda=1,\, \lambda \neq j}^{\Lambda}\pi_{\lambda}(\mathrm{d}x^{\lambda}). 
		\end{align*}
	\end{proposition} 
	The proof of Proposition \ref{prop:anytime:exch} is given in Appendix \ref{app:proof:prop}.	
	Conditioning on $x^j$, $j$ and $l$ we obtain
	\begin{equation}
		\label{eq:anytime:cond}
		A(\text{d}x^{1:\Lambda\setminus	j}\, | \, x^j, l, j) = \prod_{\lambda=1,\, \lambda \neq j}^{\Lambda}\pi_{\lambda}(\text{d}x^{\lambda}).
	\end{equation}	
	Therefore, if exchange moves on the conditional $A(\text{d}x^{1:\Lambda\setminus	j}\, | \, x^j, l, j)$ are performed by `eliminating' the $j$-th chain to obtain the expression in \eqref{eq:anytime:cond}, they are being performed involving only chains distributed according to their respective targets $\pi_\lambda$ and thus the bias is eliminated.
	\subsection{Implementation} \label{sec:APTMC:implementation}
\begin{figure*}[htbp]
		\includegraphics[width=\textwidth]{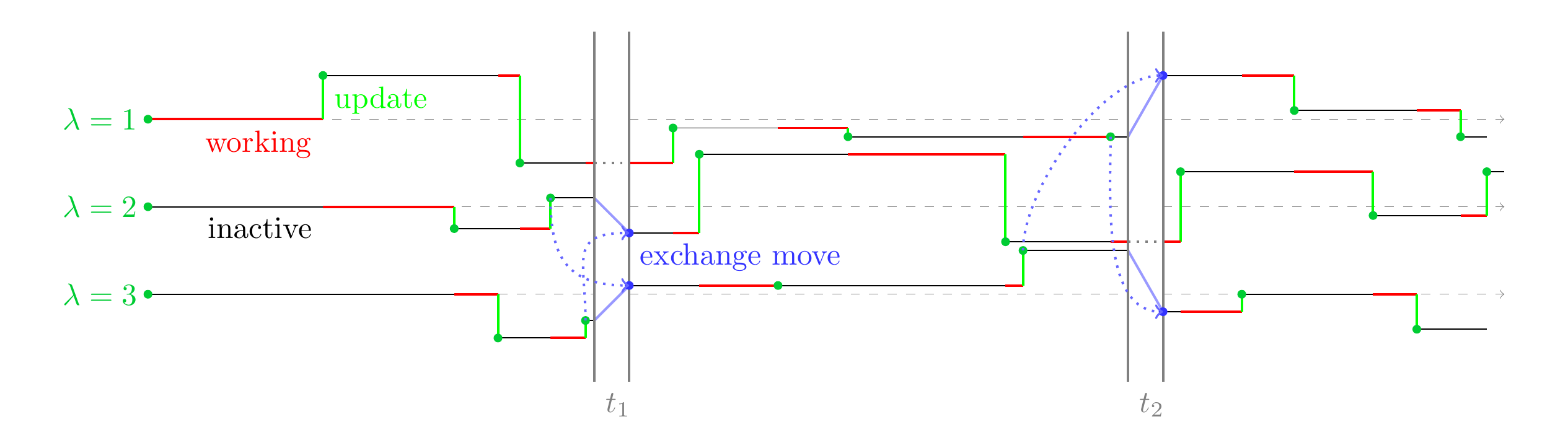}
		\caption[Progression of three chains in the Anytime Parallel Tempering Monte Carlo algorithm on a single processor.]{Illustration of the progression of three chains in the APTMC algorithm on a single processor. The \textit{green} (local move) and \textit{blue} (exchange move) dots represent samples from the posterior being recorded as their respective local and exchange moves are completed. When exchange moves occur at $t_1$, chain $\lambda=1$ is currently moving and cannot participate in exchange moves without introducing a bias. Therefore it is ignored, and the exchange moves are performed on the remaining (inactive) chains. Similarly, at time $t_2$ chain $\lambda=2$ is excluded from the exchange. The widths of intervals $t_1$ and $t_2$ are for illustrating the exchange procedure only.}
		\label{pic:APTMC:one}
	\end{figure*}

	On a single processor, the algorithm may proceed as in Algorithm \ref{alg:APTMC:one}, where in Step \ref{APTMC:one:update} the $\Lambda$ chains are simulated one at a time in a serial schedule. Figure \ref{pic:APTMC:one} provides an illustration of how the algorithm works.
	
	\begin{algorithm}[htbp]
		\caption{Anytime Parallel Tempering Monte Carlo on one processor (\texttt{APTMC-1})}\label{alg:APTMC:one}
		\begin{algorithmic}[1]
			\State Initialise real-time Markov jump process $\left(X^{1:\Lambda}, L, J\right)(0) = \left(x_0^{1:\Lambda}, 0, 1\right)$. \label{alg:APTMC:one:1}
			\State Set $n^{1:\Lambda}:=0$. \Comment{\textit{number of samples per chain}}
			\For{$i=1, 2, \ldots$} 
			\Statex \textsc{Simulate real-time Markov jump process $\left(X^{1:\Lambda}, L, J\right)(t)$ until real time $t_i$.} \label{APTMC:one:update}
			\State Perform local moves on $x^j_{n^j}$.
			\State $j := j+1 \mod \Lambda$
			\State $n^j :=  n^j+1$ \label{alg:APTMC:one:6}
			\Statex \textsc{Perform exchange steps on the conditional in \eqref{eq:anytime:cond}.}
			\State Select one or more pair(s) of adjacent chains with indices taken from the set $\{1:\Lambda\}\setminus j$.
			\State Propose to swap the selected pair(s) of states $(x^\lambda_{n^\lambda}, x^{\lambda'}_{n^{\lambda'}})$ according to Algorithm \ref{alg:APTMC:swap}.
			\EndFor
		\end{algorithmic}
	\end{algorithm}
	
	\begin{algorithm}[htbp]
		\caption{Exchange move between two chains}\label{alg:APTMC:swap}
		\begin{algorithmic}[1]
			\Statex \textbf{Input}: states $(x^\lambda_{n}, x^{\lambda'}_{n^{'}})$ where $x^\lambda_{n} \sim \pi_\lambda$ and $x^{\lambda'}_{n'}\sim\pi_{\lambda'}$.
			\State Compute exchange move acceptance probability $a_{swap}(x^\lambda_{n}, x^{\lambda'}_{n^{'}})$ given in \eqref{eq:swap}.
			\State Sample $u \sim \text{Uniform}(0, 1)$.
			\If{$u < a_{swap}(x^\lambda_{n}, x^{\lambda'}_{n^{'}})$} 
			\State $(x^\lambda_{n+1}, x^{\lambda'}_{n^{'}+1}) = (x^{\lambda'}_{n'}, x^{\lambda}_{n})$
			\Else
			\State $(x^\lambda_{n+1}, x^{\lambda'}_{n^{'}+1}) = (x^\lambda_{n}, x^{\lambda'}_{n^{'}})$
			\EndIf
			\State $n :=  n+1$ and $n^{'} := n^{'} +1$.		 
			\Statex \textbf{Output}: updated states $(x^\lambda_{n+1}, x^{\lambda'}_{n^{'}+1})$.
		\end{algorithmic}
	\end{algorithm}
	
	When multiple processors are available, the $\Lambda$ chains can be allocated to them. However, running a single chain on each processor means that when the real-time deadline occurs, all chains will be distributed according to their respective anytime distributions $\alpha_{\lambda}$, and thus be biased as exchange moves occur. Therefore, all processors must contain at least two chains. A typical scenario would be each processor is allocated two or more temperatures to sample from. 
	The implementation is defined as described in Algorithm \ref{alg:APTMC:mult}. Note that the multiple chain construction eliminates the intractable densities in the acceptance ratio for the exchange step when $\tau$ differs between processors, since exchange moves are performed between chains that are not currently working (i.e. on density \eqref{eq:anytime:cond} for a single processor and \eqref{eq:APTMC:multi} for multiple processors), so the hold time distribution does not factor in. \par 

	Depending on the problem at hand and computing resources available, there are various approaches to distributing the chains across workers. We distinguish three possible scenarios. The first is an ideal scenario, where the number of processors exceeds $\Lambda$ and the communication overhead between workers is negligible. In this scenario, each worker implements $K=2$ chains running at the same temperature. For example, with $W = \Lambda$ workers, worker $w = \lambda$ contains $2$ chains targeting $\pi_{\lambda}$. The second scenario arises when the number of workers available is limited, but communication overhead is still negligible. In this case, the chains, sorted in increasing order of temperature, are divided evenly among workers. For example, with $W = \frac{\Lambda}{2}$ workers, worker $w$ could contain two chains, one with target $\pi_{2w-1}$ and one with target $\pi_{2w}$. 
	The third scenario deals with non-negligible inter-processor communication overhead (which only affects the exchange moves). To account for this, exchange moves are divided into two types:
	\begin{enumerate}
		\item \textit{Within-worker} exchange move: performed on each individual worker in parallel, between a pair of adjacent chains. No communication between workers is necessary in this case.
		\item \textit{Between-worker} exchange move: performed by selecting a pair of adjacent workers and exchanging between the warmest eligible chain from the first worker and coldest from the second. Thus, an exchange move between two adjacent chains is effectively being performed, except this time communication between workers is required.
	\end{enumerate}

	\begin{algorithm}[htbp]
		\caption{Anytime Parallel Tempering Monte Carlo on multiple processors (\texttt{APTMC-W})}\label{alg:APTMC:mult}
		\begin{algorithmic}[1]
			\State On worker $w$, initialise the real-time Markov jump process $\left(X_w^{1:K}, L_w, J_w\right)(0) = \left(x_{w, 0}^{1:K}, 0, 1\right)$. 
			\State Set $n^{1:K}_{w} := 0$. \Comment{\textit{number of samples per chain}}
			\For{$i=1, 2, \ldots$} 
			\Statex \textsc{On each worker $w$, simulate the real-time Markov jump process $\left(X_w^{1:K}, L_w, J_w\right)(t)$ until real time $t_i$.}			
			\State Perform local moves on $x^{j_w}_{w, n_w^{j_w}}$.
			\State  $j_w := j_w+1 \mod K$
			\State $n_w^{j_w} :=  n_w^{j_w}+1$			
			\Statex \textsc{Across all workers, perform exchange steps on the conditional} 
			\begin{equation}
				\label{eq:APTMC:multi}
				A(\textbf{d}\bm{x}^{1:K\setminus\bm{j}} \, | \, \bm{x}^{\bm{j}}, \bm{l}, \bm{j}) = \prod_{w=1}^{W} \prod_{k=1,\, k \neq j_w}^{K}\pi_{w}(\text{d}x_w^{k}),
			\end{equation}
			\Statex \textsc{where} $\textbf{d}\bm{x}^{1:K\setminus\bm{j}} = \left(\text{d}x_1^{1:K\setminus	j_1}, \ldots, \text{d}x_W^{1:K\setminus	j_W}\right)$, $\bm{x}^{\bm{j}} = \left(x_1^{j_1}, \ldots, x_W^{j_W}\right)$, $\bm{l} = l_{1:W}$ \textsc{and} $\bm{j} = j_{1:W}$.
			\State For the exchange moves, combine all chains by relabelling the state indices as follows:
			$$z^{l(w, k)}_{m^{l(w, k)}} = x^{l(w, k)}_{w, n_w^{l(w, k)}},$$ where $l(w, k) = (w-1)K+l$ for $k=1, \ldots K$ and $w=1, \ldots, W$. 
			\State Select one or more pair(s) of adjacent chains with indices taken from the set $\{1:\Lambda\}\setminus l(1:W, \bm{j})$.
			\State Propose to swap the selected pair(s) of states $(z^\lambda_{m^\lambda}, z^{\lambda'}_{m^{\lambda'}})$ according to Algorithm \ref{alg:APTMC:swap}.
			\EndFor
		\end{algorithmic}
	\end{algorithm}

\subsection{Tuning considerations} \label{sec:APTMC:tuning}

In this section we discuss the issue of tuning Anytime parallel tempering by drawing on various ideas from the literature. The main concerns are the selection of the number of chains and their temperatures, the tuning of the local moves for each chain and the selection of appropriate hard deadlines for the exchange moves to occur.
In our setting, the computational budget determines the number of chains $\Lambda$, and for such a fixed budget we aim to improve sampling of the cold chain through the adoption of parallel tempering stages. The issue of determining the temperature of adjacent chains has been considered in \cite{rathore2005optimal, kone2005selection, atchade2011towards} where it was shown that an exchange success rate of approximately 20-25\% for adjacent chains is optimal, in an appropriate sense, and is demonstrated to confer the most benefit to sampling the coldest chain. However, the optimality curve (\cite{kone2005selection, atchade2011towards}) has a broad mode, and even 40\% seems appropriate. To achieve this 25\% acceptance rate of exchange moves, other than employing pilot runs, adaptive tuning is possible and \cite{miasojedow2013adaptive} use a Robbins-Munro scheme to adjust the temperatures to target a 25\% acceptance rate during runtime. The next issue is local proposals, and how large a change of state one should attempt (for the local accept/reject step). This subject has received ample attention in the literature following the seminal paper by \cite{roberts2001optimal}, where a 25\% local move acceptance rate is again optimal. The local proposal can be a Gaussian proposal whose mean and covariance matrix are tuned online (\cite{miasojedow2013adaptive}) via a Robbins-Munro scheme to achieve the 25\% local move acceptance rate. The tuning of Gaussian proposals for MCMC in general was popularised by the seminal paper of \cite{haario2001adaptive}.

When performing exchange moves, rather than selecting a single pair of adjacent chains from $\left\{(1, 2), (2, 3), \ldots, (\Lambda-1, \Lambda)\right\}$ for an exchange, it is common to propose to swap multiple pairs of chains simultaneously, as the exchange move is relatively cheap. To avoid selecting the same chain twice, they are divided into odd $\left\{(1, 2), (3, 4), \ldots\right\}$ and even $\left\{(2, 3), (4, 5), \ldots\right\}$ pairs of indices in \cite{lingenheil2009efficiency}, and all odd or even pairs are selected for exchange with equal probability. It is however shown in \cite{syed2019non} that it is better to deterministically cycle between exchanging odd and even pairs.

Although thus far we have suggested tuning the number of chains and annealing schedule for APTMC as if one were tuning a standard parallel tempering algorithm, there are some caveats which we now highlight. Selecting chains for exchange moves can be applied by omitting the currently working chains and relabelling the indices of the remaining, inactive or \textit{eligible} chains. However, note that by the nature of the Anytime exchange moves, the Anytime version of an optimised parallel tempering algorithm can be suboptimal, since one or more temperature(s) might be missing from exchange moves. Considering the example in Figure \ref{pic:APTMC:one} and assuming the chains are all running at increasing temperatures, at $t_2$, chain $2$ is working, so the exchange move is performed between chains $1$ and $3$. In a practical example, these chains would be further apart, which would lead to a lower exchange move acceptance rate. Selecting adjacent chains to target a slightly higher successful exchange rate, say 40\%, would mitigate this issue; noting that even 40\% is close to optimal  \cite{kone2005selection, atchade2011towards}. In our implementation, we only experienced a small drop in acceptance rate due to attempting to swap two eligible chains that are not immediately adjacent, and this event becomes less likely as the number of chains increases. \par

Another important facet of tuning APTMC is the issue of determining the real-time schedule $t_1, t_2, \ldots$ of exchange moves. Let $\delta$ be the real-time interval or deadline between exchange moves, so that $t_i = i\delta$ for $i=1, 2, \ldots$ We now present guidelines for calibrating $\delta$. Let $K$ be the number of chains, labelled $k=1, \ldots, K$, on the slowest processor $w_s$ (generally the one containing the cold chain), our experiments have shown that exchange moves should occur once every chain on this processor has completed at least one local move (see  \cite{dupuis2012infinite} for an alternative view advocating  an \emph{infinite} exchange frequency version of parallel tempering).
The expected hold time of one set of local moves on processor $w_s$, denoted $\mathsf{H} := \sum_{k=1}^K\mathbb{E}[H_{k}]$, can be estimated by repeatedly measuring the time taken for one set of local moves to complete, and averaging across all measurements. Using a pilot run, an estimate $\widehat{\mathsf{H}}$ of this expected hold time can be obtained, then set $\delta = \widehat{\mathsf{H}}$ for running the APTMC algorithm. This $\delta$ value can also be calibrated in real time, denoted $\delta(t)$ where $t$ is the real time. At $t=0$, initialise $\delta(0) = \delta_0$ such that $\delta_0>0$ is an initial, user-defined guess. Similarly as before, record a hold time sample every time a set of local move occurs on processor $w_s$, then after every exchange move, recompute $\widehat{\mathsf{H}}$ and update $\delta(t) = \widehat{\mathsf{H}}$. An advantage of this second approach is that $\delta(t)$ then adapts to a potentially time-inhomogeneous hold time, due e.g. to competing jobs on the processors starting mid-algorithm and suddenly slowing down the computation time of local moves. \par

A scenario we encountered in our experiments was non-negligible communication overhead between workers to execute the exchange moves, and this overhead was comparable to local move times which were themselves lengthy. To mitigate the communication overhead, as  described in Section \ref{sec:APTMC:implementation}, exchange moves are divided into within- and between- workers.
On a given worker with $K$ chains, a set of worker specific moves is performed before inter-worker exchanges. These were $K$ local moves, one (set of) within-worker exchange moves, then $K$ more local moves, before inter-worker communication occurs for between-worker exchange moves. Given that within-worker exchanges are instant, this amounts in real time to performing $2K$ local moves on this worker before inter-worker communication occurs. Therefore, the real-time deadline in the Anytime version of the algorithm for this scenario is set to be $\delta = 2\mathsf{H}$ and can be determined as above. See Section \ref{settings:LV:multi} for an example.\par
Finally, Section \ref{sec:ABC:MCMC:MA} details other, empirical tools that help with tuning by assessing the efficiency of each chain. These include evaluating the sample autocorrelation function (acf), as well as the integrated autocorrelation time ($IAT$) and effective sample size ($ESS$).

	\section{Application to Approximate Bayesian Computation (ABC)} \label{cha:APTMC:ABC}
In this section we adapt the APTMC framework to ABC.

\subsection{Overview of ABC} \label{sec:APTMC:ABC:ov}

The notion of ABC was developed by \cite{tavare1997inferring} and \cite{pritchard1999population}. It can be seen as a likelihood-free way to perform Bayesian inference, using instead simulations from the model or system of interest, and comparing them to the observations available.\par

Let $y \in \mathbb{R}^d$ be some data with underlying unknown parameters $\theta \sim p(\text{d}\theta)$, where $p(\theta)$ denotes the prior for $\theta \in \Theta$. Suppose we are in the situation in which the likelihood $f(y \, | \, \theta)$ is either intractable or too computationally expensive, which means that MCMC cannot be performed as normal. Assuming that it is possible to sample from the density $f(\, \cdot \, | \, \theta)$ for all $\theta \in \Theta$, approximate the likelihood by introducing an artificial likelihood $f^{\varepsilon}$ of the form 
\begin{equation}
	f^{\varepsilon}(y \, | \, \theta) = \text{Vol}(\varepsilon)^{-1} \int_{B_{\varepsilon}(y)} f(x\,|\,\theta) \text{d}x,
\end{equation} 
where $B_{\varepsilon}(y)$ denotes a metric ball centred at $y$ of radius $\varepsilon > 0$ and Vol$(\varepsilon)$ is its volume. The resulting approximate posterior is given by
\begin{equation*}
	p^{\varepsilon}(\theta\,|\,y) = \frac{p(\theta) f^{\varepsilon}(y\,|\,\theta)}{\int p(\vartheta) f^{\varepsilon}(y\,|\,\vartheta) \text{d}\vartheta}.
\end{equation*}
The likelihood $f^{\varepsilon}(y\,|\,\theta)$ cannot be evaluated either, but an MCMC kernel can be constructed to obtain samples from the approximate posterior $\pi^{\varepsilon}(\theta, x)$ defined as
\begin{align*}
	\pi^{\varepsilon}(\theta, x) &= p^{\varepsilon}(\theta, x \,|\,y)
	\propto p(\theta)f(x| \theta) \mathbbm{1}_{\varepsilon}(x) \text{Vol}(\varepsilon)^{-1},
\end{align*}
where $\mathbbm{1}_{\varepsilon}(x)$ is the indicator function for $x \in B_{\varepsilon}(y)$. This is referred to as \textit{hitting} the ball $B_{\varepsilon}(y)$. In the MCMC kernel, one can propose $\theta' \sim q(\text{d}\theta' \,|\,\theta)$ for some proposal density $q$, simulate the dataset $x \sim f(\text{d}x \,|\,\theta')$ and accept $\theta'$ as a sample from the posterior if $x \in B_{\varepsilon}(y)$. \par

The \textit{1-hit MCMC kernel}, proposed by \cite{lee2012choice} and described in Algorithm \ref{alg:ABC:1hit} introduces local moves in the form of a `race': given current and proposed parameters $\theta$ and $\theta'$, respectively simulate corresponding datasets $x$ and $x'$ sequentially. The state associated with the first dataset to hit the ball $B_{\varepsilon}(y)$ `wins' and is accepted as the next sample in the Markov chain. The proposal $\theta'$ is also accepted if both $x$ and $x'$ hit the ball at the same time. 

\begin{algorithm}[htbp]
	\caption{1-hit MCMC kernel for ABC}\label{alg:ABC:1hit}
	\begin{algorithmic}[1]
		\Statex \textbf{Input}: current state $\left(\theta_n, x_n\right)$.
		\State Propose $\theta' \sim q(\text{d}\theta \,|\,\theta_n)$. \Comment{\textit{propose a local move}}
		\State Compute preliminary acceptance probability. \Comment{\textit{prior check}}
		$$a(\theta_n, \theta') = \min\left\{1, \frac{p(\theta') q(\theta_n \,|\,\theta')}{p(\theta_n) q(\theta' \,|\,\theta_n)}\right\}.$$
		\State Sample $u \sim \text{Uniform}(0, 1)$ .
		\If{$u < a(\theta_n, \theta')$} 
		\State \textsc{race} $:=$ \textsc{true}
		\Else
		\State \textsc{race} $:=$ \textsc{false} 
		\State Retain $\left(\theta_{n+1}, x_{n+1}\right) = \left(\theta_n, x_n\right)$. \Comment{\textit{reject $\theta'$ as it is unlikely to win race}}
		\EndIf
		\While{\textsc{race}} 	
		\State Simulate $x \sim f(\text{d}x \, | \,\theta_n)$ and $x' \sim f(\text{d}x' \, |\, \theta')$.
		\If{$x \in B_{\varepsilon}\left(y\right)$ \textbf{or} $x' \in B_{\varepsilon}\left(y\right)$} \Comment{\textit{stop the race once either $x$ or $x'$ hits the ball}}
		\State \textsc{race} $:=$ \textsc{false}
		\EndIf
		\EndWhile
		\If {$x' \in B_{\varepsilon}\left(y\right)$} \Comment{\textit{accept or reject move}}
		\State Set $\left(\theta_{n+1}, x_{n+1}\right) = \left(\theta', x'\right)$.
		\Else
		\State Retain $\left(\theta_{n+1}, x_{n+1}\right) = \left(\theta_n, x\right)$. 
		\EndIf
		\Statex \textbf{Output}: updated state $\left(\theta_{n+1}, x_{n+1}\right)$.
	\end{algorithmic}
\end{algorithm}

\subsection[ABC-APTMC]{Anytime Parallel Tempering Monte Carlo for Approximate Bayesian Computation (ABC-APTMC)} \label{sec:ABC:APTMC}

Including the 1-hit kernel in the local moves of a parallel tempering algorithm is straightforward. Exchange moves must however be adapted to this new likelihood-free setting. Additionally, the race that occurs takes a random real time to complete, and this time is temperature-dependent, as it is quicker to hit a ball of larger radius. This provides good motivation for the use of Anytime Monte Carlo.

\subsubsection{Exchange moves}
The exchange moves for ABC are derived similarly as in \cite{baragatti2013likelihood}. Let $(\theta, x)$ and $(\theta', x')$ be the states of two chains targeting $\pi^{\varepsilon}$ and $\pi^{\varepsilon'}$, respectively, where $\varepsilon'>\varepsilon$. Here, this is equivalent to saying $\theta'$ is the state of the `warmer' chain. We already know that $x'$ falls within $\varepsilon'$ of the observations $y$, i.e. $x' \in B_{\varepsilon'}(y)$. Similarly, we also know that $x \in B_{\varepsilon}(y)$, and trivially that $x \in B_{\varepsilon'}(y)$. If $x'$ also falls within $\varepsilon$ of $y$, then swap the states, otherwise do not swap. The odds ratio is
\begin{align*}
	\frac{\pi^{\varepsilon'}(\theta, x)\pi^{\varepsilon}(\theta', x')}{\pi^{\varepsilon}(\theta, x)\pi^{\varepsilon'}(\theta', x')}
	= \frac{p(\theta)f(x \,|\, \theta)\text{Vol}(\varepsilon')p(\theta')f(x' \,|\, \theta')\mathbbm{1}_{\varepsilon}(x')\text{Vol}(\varepsilon)}
	{p(\theta)f(x \,|\, \theta)\text{Vol}(\varepsilon)p(\theta')f(x' \,|\, \theta')\text{Vol}(\varepsilon')}
	= \mathbbm{1}_{\varepsilon}(x'),
\end{align*}
so the probability of the swap being accepted is the probability of $x'$ also hitting the ball of radius $\varepsilon$ centred at $y$. This type of exchange move is summarised in Algorithm \ref{alg:ABC:exch:fast}. 

\subsubsection{Implementation}
The full implementation of the ABC-APTMC  algorithm on a single processor (\texttt{ABC-APTMC-1}) is described in Algorithm \ref{alg:ABC:APTMC}. The multi-processor algorithm can similarly be modified to reflect these new exchange moves and the resulting algorithm is referred to as \texttt{ABC-APTMC-W}.
\begin{algorithm}[htbp]
	\caption{ABC: exchange move between two chains}\label{alg:ABC:exch:fast}
	\begin{algorithmic}[1]
		\Statex \textbf{Input}: states $\omega_n = \left((\theta, x), (\theta', x')\right)$ where $\theta \sim \pi$, $x \sim f(\text{d}x \, |\, \theta)$ and $\theta' \sim \pi'$, $x' \sim f(\text{d}x' \, |\, \theta')$.
		\Statex \Comment{\textit{both $(\theta, x)$ and $(\theta', x')$ are outputs from Algorithm \ref{alg:ABC:1hit} for different $\varepsilon' > \varepsilon$}}
		\If {$x' \in B_{\varepsilon}\left(y\right)$} \Comment{\textit{accept or reject swap depending on whether $x'$ also hits the ball of radius $\varepsilon$}}
		\State Set $\omega_{n+1} = \left((\theta', x'), (\theta, x)\right)$.
		\Else
		\State Retain $\omega_{n+1} = \omega_n$.
		\EndIf
		\State $n := n+1$
		\Statex \textbf{Output}: updated states $\omega_{n+1}$.
	\end{algorithmic}
\end{algorithm}

\begin{algorithm}[htbp]
	\caption{ABC: Anytime Parallel Tempering Monte Carlo Algorithm (\texttt{ABC-APTMC-1})}\label{alg:ABC:APTMC}
	\begin{algorithmic}[1]
		\State Initialise the real-time Markov jump process $(\theta^{1:\Lambda}, L, J) = (\theta_0^{1:\Lambda}, 0, 1)$. \label{alg:ABC:APTMC:init}
		\State Set $n:=0$.
		\For{$i:=1, 2, \ldots$}
		\Statex \textsc{Simulate the real-time Markov jump process} $(\theta, L, J)(t)$ \textsc{until real time} $t_i$. 
		\State Perform local moves on $\left(\theta_n^{j}, x_n^j\right)$ according to Algorithm \ref{alg:ABC:1hit}.
		\State  $j := j+1 \mod \Lambda$
		\Statex \textsc{Perform exchange steps on the conditional}:
		$$A(\text{d}\theta^{1:\Lambda} \,|\,\theta^j, l, j) = \prod_{\lambda=1, \lambda \neq j}^{\Lambda} \pi_{\lambda}(\text{d}\theta^{\lambda}).$$
		\State Perform exchange moves on $\omega_n = \left((\theta_n^{\lambda}, x_n^{\lambda}), (\theta_n^{\lambda'}, x_n^{\lambda'})\right)$ according to Algorithm \ref{alg:ABC:exch:fast}.
		\EndFor
	\end{algorithmic}
\end{algorithm}

\section{Experiments}\label{cha:exp}

In this section, we first illustrate the workings of the algorithms presented in Section \ref{sec:APTMC:implementation} on a simple model, in which real-time behaviour is simulated using virtual time and an artificial hold distribution. The model is also employed to demonstrate the gain in efficiency provided by the inclusion of exchange moves. Then, the ABC version of the algorithms, as presented in Section \ref{cha:APTMC:ABC}, is applied to two case studies. The first case is a simple model and serves to verify the workings of the ABC algorithm, including bias correction. The second case considers the problem of estimating the parameters of a stochastic Lotka-Volterra predator-prey model $-$ in which the likelihood is unavailable $-$ and serves to evaluate the performance of the Anytime parallel tempering version of the ABC-MCMC algorithm, as opposed to the standard versions (with and without exchange moves) on both a single and multiple processors. The exchange moves are set up so that multiple pairs could be swapped at each iteration. All experiments in this paper were run on \textsc{Matlab} and the code is available at \url{https://github.com/alixma/ABCAPTMC.git}.

\subsection{Analytic example: Gamma mixture model} \label{cha:APTMC:GM}

In this example we attempt to sample from an equal mixture of two Gamma distributions using the APTMC algorithm. Define the target $\pi(\text{d}x)$ and an `artificial' hold time $\tau(\text{d}h \, | \, x)$ distribution as follows:
\begin{align*}
	X &\sim \phi \, \text{Gamma}(k_1, \theta_1) + (1-\phi) \, \text{Gamma}(k_2, \theta_2), \\
	H \, | \, x &\sim \psi \, \text{Gamma}\left(\frac{x^p}{\theta_1}, \theta_1\right) + (1-\psi) \, \text{Gamma}\left(\frac{x^p}{\theta_2}, \theta_2\right),
\end{align*}
with mixture coefficients $\phi = \frac{1}{2}$ and $\psi$, where Gamma$(\,\cdot\,,\,\cdot\,)$ denotes the probability density function of a Gamma distribution, with shape and scale parameters $(k_1, \theta_1)$ and $(k_2, \theta_2)$ for each components, respectively, and with polynomial degree $p$, assuming it remains constant for both components of the mixture.\par
In the vast majority of experiments, the explicit form of the hold time distribution $\tau$ is not known, but observed in the form of the time taken by the algorithm to simulate $X$. For this example, so as to avoid external factors such as competing jobs affecting the hold time, we assume an explicit form for $\tau$ is known and simulate virtual hold times. This consists of simulating a hold time $h \sim \tau(\text{d}h \, | \, x)$ and advancing the algorithm forward for $h$ units of virtual time without updating the chains, effectively `pausing' the algorithm. These virtual hold times are introduced such that what in a real-time example would be the effects of constant $(p=0)$, linear $(p=1)$, quadratic $(p=2)$ and cubic $(p=3)$ computational complexity can be studied. Another advantage is that the anytime distribution $\alpha_{\Lambda}(\text{d}x)$ of the cold chain can be computed analytically and is the following mixture of two Gamma distributions
\begin{align} \label{eq:toy:anytime}
	\alpha_{\Lambda}(\text{d}x) &= \varphi(p, k_{1:2}, \theta_{1:2}) \, \text{Gamma}\left(k_1+p, \theta_1\right) \nonumber \\
	&+ [1 - \varphi(p, k_{1:2}, \theta_{1:2})] \, \text{Gamma}\left(k_2+p, \theta_2\right),
\end{align}
where 
\begin{equation*}
	\varphi(p, k_{1:2}, \theta_{1:2}) = \left(1 + \frac{\Gamma(k_1) \Gamma(p+k_2) \theta_2^{p}}{\Gamma(k_2) \Gamma(p+k_1) \theta_1^{p}}\right)^{-1}.
\end{equation*}
We refer the reader to Appendix \ref{app:proof:anytime} for the proof of \eqref{eq:toy:anytime}. In the anytime distribution, one of the components of the Gamma distribution will have an associated mixture coefficient $\varphi(p, k_{1:2}, \theta_{1:2})$ or $1-\varphi(p, k_{1:2}, \theta_{1:2})$  which increases with $p$ while the coefficient of the other component decreases proportionally.  Note that for constant $(p=0)$ computational complexity, the anytime distribution is equal to the target distribution $\pi$. 

\subsubsection{Implementation} \label{sec:APTMC:toy:one}
On a single processor, the Anytime Parallel Tempering Monte Carlo algorithm (referred to as \texttt{APTMC-1}) is implemented as follows: simulate $\Lambda = 8$ Markov chains, each targeting the distribution $\pi_{\lambda}(\text{d}x) = \pi(\text{d}x)^{\frac{\lambda}{\Lambda}}$. To construct a Markov chain $(X^{\lambda})_{n=0}^{\infty}$ with target distribution 
$$\pi_{\lambda}(x) \propto \left[\frac{1}{2}\text{Gamma}\left(k_1, \theta_1\right) + \frac{1}{2} \, \text{Gamma}\left(k_2, \theta_2\right)\right]^{\frac{\lambda}{\Lambda}}$$	
for $\lambda = 1, \ldots, \Lambda$, use a \textit{Random Walk Metropolis} update, i.e. symmetric Gaussian proposal distribution $\mathcal{N}(x_n^{\lambda}, \sigma^2)$ with mean $x_n^{\lambda}$ and standard deviation $\sigma = 0.5$. Set $(k_1, k_2) = (3, 20)$, $(\theta_1, \theta_2) = (0.15, 0.25)$ and use $p \in \left\{0, 1, 2, 3\right\}$. The single processor algorithm is run for $T = 10^8$ units of virtual time, with exchange moves alternating between occurring on all odd $(1, 2), (3, 4), (5, 6)$ and all even $(2, 3), (4, 5), (6, 7)$ pairs of inactive chains every $\delta = 5$ units of virtual time. When the algorithm is running, a sample is recorded every time a local or exchange move occurs.\par

On multiple processors, the APTMC algorithm (referred to as \texttt{APTMC-W}) is implemented similarly. A number of $W = \Lambda = 8$ processors is used, where each worker $w = \lambda$ contains $K=2$ chains, all targeting the same $\pi_{\lambda}$ for $\lambda = 1, \ldots, \Lambda$. The multiple processor algorithm is run for $T = 10^7$ units of virtual time, with exchange moves alternating between occurring on all odd $(1, 2), (3, 4), (5, 6), (7, 8)$ and all even $(2, 3), (4, 5), (6, 7)$ pairs of workers every $\delta=5$ units of virtual time. On each worker, the chain which was not working when calculations were interrupted is the one included in the exchange moves.

\subsubsection{Verification of bias correction}
To check that the single and multiple processor algorithms are successfully correcting for bias, they are also run \textit{uncorrected}, i.e. not excluding the currently working chain. This means that exchange moves are also performed on samples distributed according to $\alpha$ instead of $\pi$, thus causing the algorithm to yield biased results. Since the bias is introduced by the exchange moves (when they are performed on $\alpha$), we attempt to create a `worst case scenario', i.e. maximise the amount of bias present when the single processor algorithm is uncorrected. The algorithm is further adjusted such that local moves are not performed on the cold chain and it is instead solely made up of samples resulting from exchange moves with the warmer chains. 
The multi-processor \texttt{APTMC-W} algorithm is not run in a `worst case scenario', so local moves on the cold chain of the multi-processor algorithm are therefore allowed. This is meant to reveal how the bias caused by failing to correct when performing exchange moves across workers is still apparent, if less strongly. \par
\begin{figure*}[htbp]
	\centering
	\includegraphics[width=\textwidth]{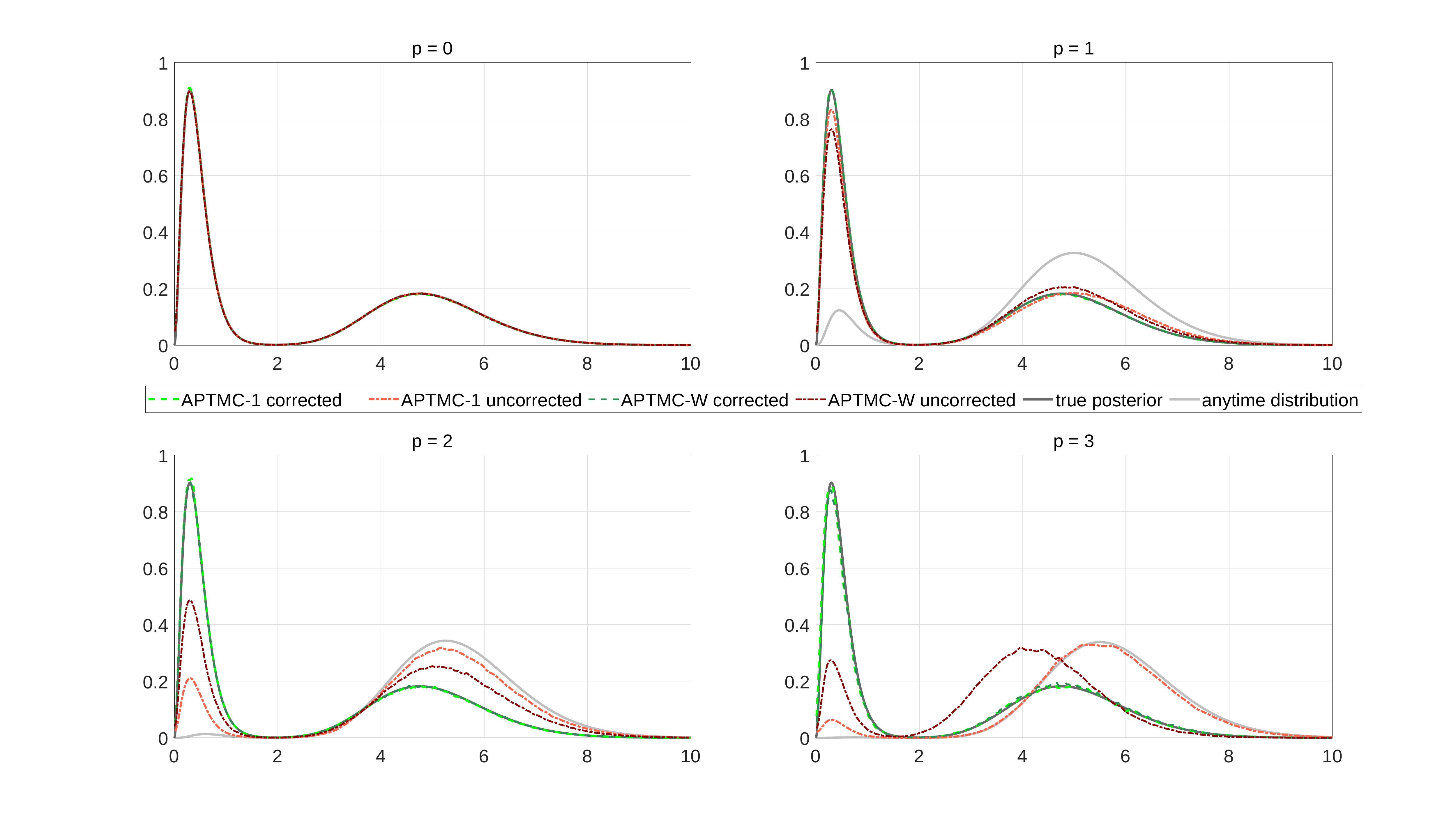}
	\caption[Density estimates of the cold chain for runs of the single and multi-processor \texttt{APTMC} algorithm]{Density estimates of the cold chain for bias corrected and uncorrected runs of the single (\texttt{APTMC-1}) and multi-processor (\texttt{APTMC-W}) algorithms on various hold time distributions $p \in \left\{0, 1, 2, 3\right\}$. In the single-processor case, the cold chain is made up entirely of updates resulting from exchange moves. The solid \textit{dark grey} line represents the true posterior density $\pi$ and the solid \textit{light grey} line the anytime distribution $\alpha$. The case $p=0$ represents an instance in which, in a real-time situation, the local moves do not take a random time to complete, and therefore all densities are identical. The two \textit{green} dashed  lines represent bias corrected densities and the \textit{red} dot-dashed represent uncorrected densities. For $p\geq1$, the two corrected densities are identical to the posterior, indicating that the bias correction was successful.}
	\label{fig:one:uncorrected}
\end{figure*}

Figure \ref{fig:one:uncorrected} shows kernel density estimates of the post burn-in cold chains resulting from runs of the \texttt{APTMC-1} and \texttt{APTMC-W} algorithms, uncorrected and corrected for bias. As expected, when the hold time does not depend on $x$, which corresponds to the case there $p=0$, no bias is observed. On the other hand, the cold chains for the single-processor algorithm with computational complexity $p \in \left\{1, 2, 3\right\}$ have been corrupted by biased samples and converged to a shifted distribution which puts more weight the second Gamma mixture component, instead of an equal weight. Additionally, the bias becomes stronger as computational complexity $p$ increases. A similar observation can be made for the cold chains from the multi-processor experiment $-$ which display a milder bias due to local moves occurring on the cold chain. The green dashed densities indicate that when the algorithms are corrected, i.e. when the currently working chain is not included in exchange moves, it successfully eliminates the bias for all $p \in \left\{1, 2, 3\right\}$ to return the correct posterior $\pi$ $-$ despite even this being the `worst case scenario' in the case of the \texttt{APTMC-1} algorithm. Note that the uncorrected density estimates do not exactly correspond to the anytime distributions. This has nothing to do with burn-in, but with the proportion of biased samples (from exchange moves) present in the chain.
\subsubsection{Performance evaluation} \label{sec:ABC:MCMC:MA}
Next we verify that introducing the parallel tempering element to the Anytime Monte Carlo algorithm improves performance. A standard \texttt{MCMC} algorithm is run for computational complexities $p \in \left\{0, 1, 2, 3\right\}$, applying the random walk Metropolis update described in Section \ref{sec:APTMC:toy:one}. The single and multiple processor APTMC algorithms are run again for the same amount of virtual time, with exchange moves occurring every $\delta_{0:2}=5$ units of virtual time for $p\leq2$ and every $\delta_3 = 30$ units when $p=3$. The single processor version is run on $\Lambda_s = 8$ chains, and the multi-processor on $W = 8$ workers, with $K=2$ chains per worker, so $\Lambda_m = 16$ chains in total. This time, local moves are performed on the cold chain of the single processor \texttt{APTMC-1} algorithm.\par 
To compare results, kernel density estimates of the posterior are obtained from the post burn-in cold chains for each algorithm using the \texttt{kde} function in \cite{MATLAB2019}, developed by \cite{botev2010kernel}. It is also important to note that even though all algorithms run for the same (virtual) duration, the standard \texttt{MCMC} algorithm is performing local moves on a single chain uninterrupted until the deadline, while the \texttt{APTMC-1} algorithm has to update $\Lambda=8$ chains in sequence, and each worker $w$ of the \texttt{APTMC-W} algorithm has to update $K=2$ chains in sequence before exchange moves occur. Therefore, by (virtual) time $T$ the algorithms will not have returned samples of similar sizes. For a fair performance comparison, the sample autocorrelation function (acf) is estimated first of all. When available, the acf is averaged over multiple chains to reduce variance in its estimates. Other tools employed are
\begin{itemize}
	\item \textit{Integrated Autocorrelation Time} ($IAT$), the computational inefficiency of an MCMC sampler. Defined as 
	\begin{equation*}
		IAT_s = 1+2\sum_{\ell=1}^{\infty} \rho_s({\ell}),
	\end{equation*}
	where $\rho_s({\ell})$ is the autocorrelation at the $\ell$-th lag of chain $s$. It measures the average number of iterations required for an independent sample to be drawn, or in other words the number of correlated samples with same variance as one independent sample. Hence, a more efficient algorithm will have lower autocorrelation values and should yield a lower $IAT$ value. Here, the $IAT$ is estimated using a method initially suggested in \cite{sokal1997monte} and \cite{goodman2010ensemble}, and implemented in the Python package \texttt{emcee} by \cite{foreman2013emcee} (Section 3). Let  
	\begin{equation*}
		\hat{IAT}_s = 1 + 2\sum_{\ell=1}^{M} \hat{\rho}_s({\ell}),
	\end{equation*}
	where $M$ is a suitably chosen cutoff, such that noise at the higher lags is reduced. Here, the smallest $M$ is chosen such that $M \geq C\hat{\rho}_s(M)$ where $C\approx6$. More information on the choice of $C$ is available in \cite{sokal1997monte}.
	\item \textit{Effective Sample Size} ($ESS$), the amount of information obtained from an MCMC sample. It is closely linked to the $IAT$ by definition:
	\begin{equation*}
		ESS_s = \frac{N_s}{IAT_s},
	\end{equation*}
	where $N_s$ is the size of the current sample $s$. The $ESS$ measures the number of independent samples obtained from MCMC output. 
\end{itemize}	
As per \cite{foreman2013emcee}, when multiple repeat runs of an experiment are performed (see Section \ref{sec:ABC:LV}), the $IAT$ for a given algorithm is obtained by averaging the acf returned by this algorithm over the repeat runs, and the resulting $ESS$s of each run are summed to obtain a cumulative $ESS$ for this algorithm.\par
The resulting $ESS$ and $IAT$ for different algorithms and computational complexities are computed and shown in Table \ref{table:mean_ESS_IAT}. 
If an exchange move is accepted, the new state of the chain does not depend on the value of the previous state. This means that the autocorrelation in a chain containing a significant proportion of (accepted) samples originating from exchange moves will be lower. For low $p$, significantly more local moves occur before each deadline, as hold times are short, while for a higher $p$, the hold times are longer and hence fewer local moves are able to occur. Therefore, higher values of $p$ will yield a higher proportion of samples from exchange moves, and thus a more notable increase in efficiency.\par

\begin{figure*}[htbp]
	\centering
	\includegraphics[width=\textwidth]{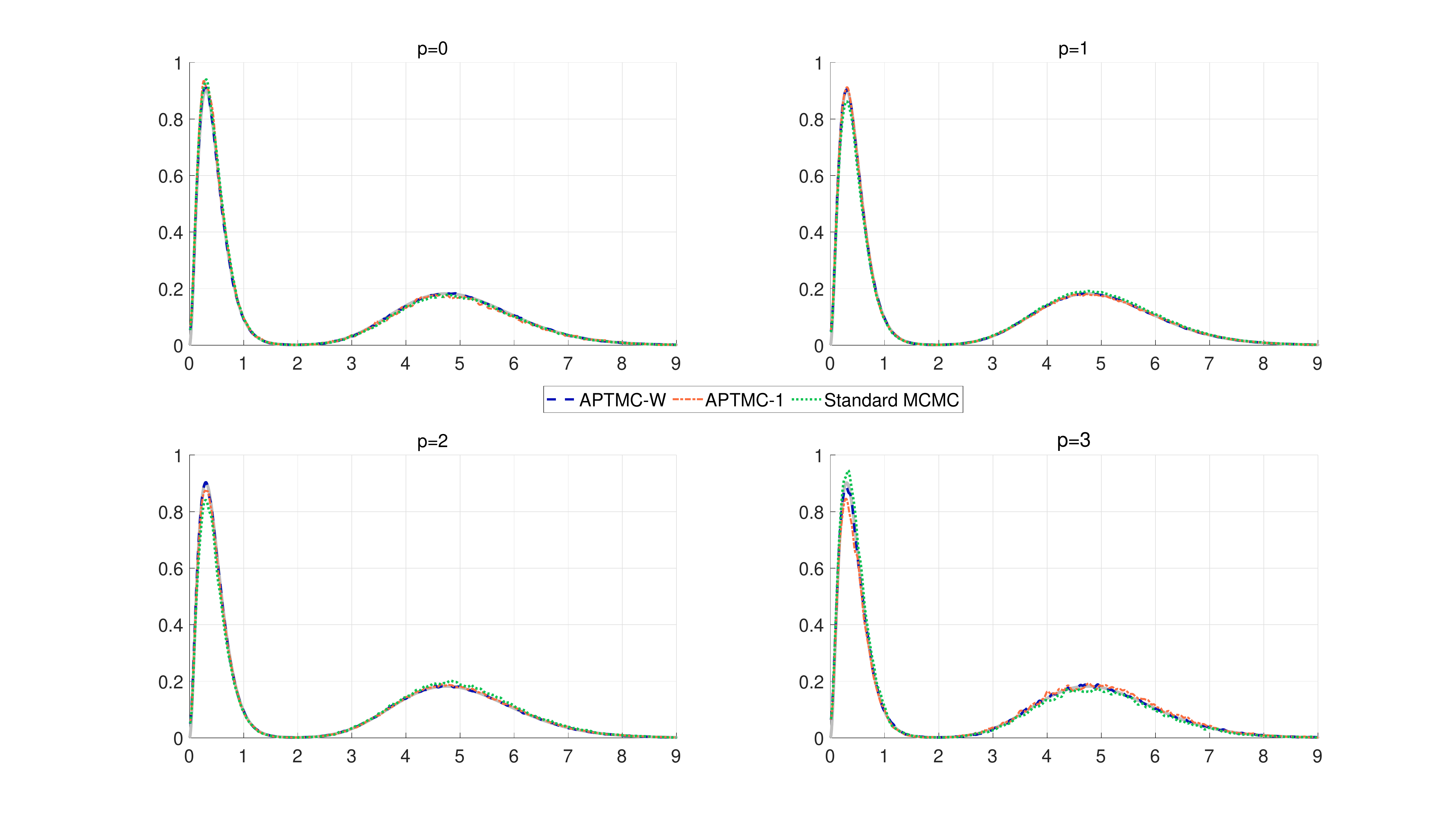}
	\caption[Density estimates of the cold posterior for runs of the single, multi-processor \texttt{APTMC} and standard (\texttt{AMC}) algorithms.]{Density estimates of the cold posterior for runs of the single (\textit{orange}) and multiple (\textit{blue}) processor APTMC algorithms (\texttt{APTMC-1} and \texttt{APTMC-W}, respectively) as well as the standard (\textit{green}) \texttt{MCMC} algorithm. The \textit{grey} line represents the true posterior density $\pi$. Each plot corresponds to a different hold time distribution $p \in \left\{0, 1, 2, 3\right\}$. While the multi processor density has successfully converged for all $p$ $-$ as evidenced by the perfect overlap between the grey and dark blue lines $-$, the other two algorithms tend to struggle more and more to estimate the first mode of the posterior as $p$ increases.}
	\label{fig:posterior:compare}
\end{figure*}

In Figure \ref{fig:posterior:compare} we observe, that the quality of the posterior estimates decreases as $p$ increases. As a matter of fact, $10^7$ units of virtual time tend to not be enough for the some of the posterior chains to completely converge. Indeed, while the standard \texttt{MCMC} algorithm performs reasonably well for $p=0$, it becomes increasingly harder for it to fully converge for higher computational complexities. Similarly, the single processor \texttt{APTMC-1} algorithm returns reasonably accurate posterior estimates for $p\leq2$ but then visibly underestimates the first mode of the true posterior for $p=3$. In general, the multi-processor \texttt{APTMC-W} algorithm returns results closest to the true cold posterior for all $p$. \par
\begin{figure*}[htbp]
	\centering
	\includegraphics[width=\textwidth]{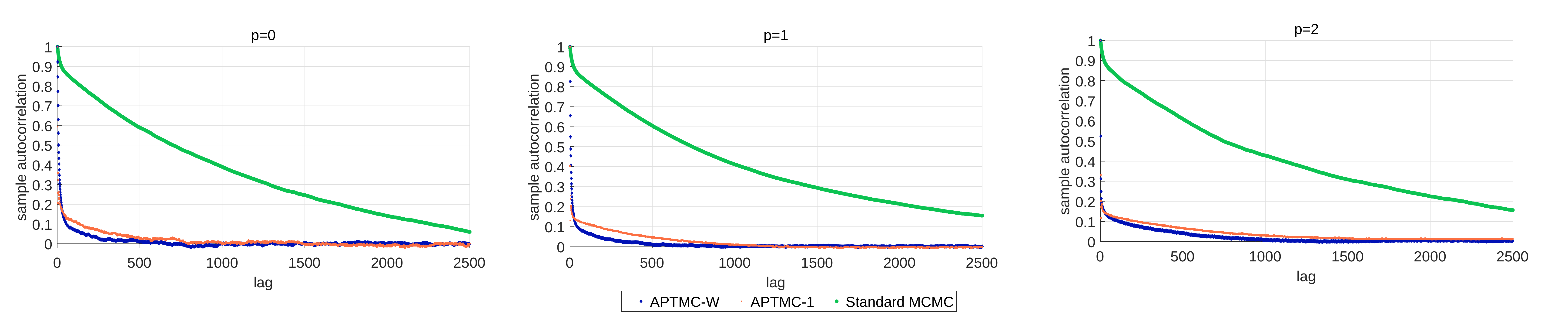}
	\caption[Sample acf of the cold chain for runs of the single, multi-processor \texttt{APTMC} and standard \texttt{AMC} algorithms]{Plots of the sample autocorrelation function up to lag $2500$ of the post burn-in cold chain for runs of the single (\textit{orange}) and multiple (\textit{blue}) processor APTMC algorithms (\texttt{APTMC-1} and \texttt{APTMC-W}, respectively) as well as for the output of the standard Anytime Monte Carlo (\texttt{MCMC}) algorithm (\textit{green}). Each plot corresponds to a different computational complexity $p \in \left\{0, 1, 2\right\}$. The two \texttt{APTMC} algorithms perform considerably better than standard \texttt{MCMC} for all $p$. The sample acf plot for $p=3$ has been omitted due to both the \texttt{APTMC-1} and \texttt{MCMC} chains not having fully converged to their posterior.}
	\label{fig:acf:compare}
\end{figure*}

	
As for efficiency, Table \ref{table:mean_ESS_IAT} displays a much lower $IAT$ and much higher $ESS$ for both \texttt{APTMC} algorithms, indicating that they are much more efficient than the standard \texttt{MCMC} algorithm. This is further supported by the sample autocorrelation decaying much more quickly for \texttt{APTMC} algorithms than for the \texttt{MCMC} algorithm for all $p$ in Figure \ref{fig:acf:compare}. The multi-processor \texttt{APTMC-W} algorithm also yields $IAT$ values that are lower than those returned by the single processor \texttt{APTMC-1} algorithm for $p<3$, and similarly yields $ESS$s that are higher for all $p$. The $ESS$ and $IAT$ values for chains that have not converged to their posterior (their resulting kernel density estimates significantly under or overestimate modes in Figure \ref{fig:posterior:compare}) have been omitted from the table.\par 
Next, we consider an application of the APTMC framework to ABC, a class of algorithms that are well-adapted to situations in which the likelihood is either intractable or computationally prohibitive. ABC features a real hold time at each MCMC iteration, making it an ideal candidate for adaptation to the Anytime parallel tempering framework.
\begin{table}[htbp]                                          
	\centering                                              
	\begin{tabular}{c cc cc cc}  
		\toprule
		$p$ & \multicolumn{2}{c}{Multi-processor}	& \multicolumn{2}{c}{Single-processor}	& \multicolumn{2}{c}{Standard}	\\           
		& \multicolumn{2}{c}{\texttt{APTMC-W}}	& \multicolumn{2}{c}{\texttt{APTMC-1}}	& \multicolumn{2}{c}{\texttt{MCMC}}	\\      
		& 	$IAT$ 	&	$ESS$ 	& 	$IAT$ 	&	$ESS$ 	& 	$IAT$ 	&		$ESS$				\\ 																							
		\midrule                                                 
		$0$ & 	53.925	&	12049	& 	81.156  &	1202.2 	& 	1739.0	&	287.46			\\ 
		$1$ & 	45.942	&	5888.3 	& 	95.104  &	708.74 	& 	2818.2	&	64.047			\\ 
		$2$ & 	80.871	&	1168.4	& 	132.79 	&	448.92  & - & - \\
		$3$ & 	131.91	&	116.51	& 	-	& 	- 	& 	-	&	-			\\ 
		\bottomrule                                                  
	\end{tabular}                                           
	\caption[$IAT$ and $ESS$ for runs of the Anytime single, multi-processor \texttt{APTMC} and standard \texttt{MCMC} algorithms]{Integrated autocorrelation time ($IAT$) and effective sample size ($ESS$) for runs of the single, multi-processor Anytime parallel tempering and standard MCMC algorithms. The algorithms were run for $10^6$ units of virtual time for computational complexity $p=0$ and $10^7$ units for $p\geq 1$, and the resulting $ESSs$ were scaled down for consistency with $p=0$. The $ESS$ and $IAT$ values for chains that have not converged to their posterior (their resulting kernel density estimates significantly under or overestimate modes in Figure \ref{fig:posterior:compare}) have been omitted.}
	\label{table:mean_ESS_IAT}                             
\end{table} 


\subsection{ABC example: univariate Normal distribution} \label{sec:ABC:MCMC:unormal}
To validate the results of Section \ref{sec:ABC:APTMC}, consider another simple example, initially featured in \cite{lee2012choice}, and adapted here within the APTMC framework. Let $Y$ be a Gaussian  random variable, i.e. $Y \sim \mathcal{N}(\theta, \sigma^2)$, where the standard deviation $\sigma$ is known but the mean $\theta$ is not. The ABC likelihood here is 
$$f^{\varepsilon}(y \,|\,\theta) = \Phi\left(\frac{y +\varepsilon - \theta}{\sigma}\right) - \Phi\left(\frac{y -\varepsilon - \theta}{\sigma}\right)$$ for $\varepsilon > 0$ where $\Phi(\,\cdot\,)$ is the cumulative distribution function (cdf) of a standard Gaussian. Using numerical integration tools in \textsc{Matlab}, it is possible to obtain a good approximation of the true posterior for any $\varepsilon$ for visualisation. Let $y=3$ be an observation of $Y$ and $\sigma^2=1$, and put the prior $p(\theta) = \mathcal{N}\left(0, 5\right)$ on $\theta$. In this example, the exact posterior distribution for $\theta$ can straightforwardly be shown to be $\mathcal{N}\left(\frac{5}{2}, \frac{5}{6}\right)$.\par
\begin{figure*}[htbp]
	\centering
	\includegraphics[width=\textwidth]{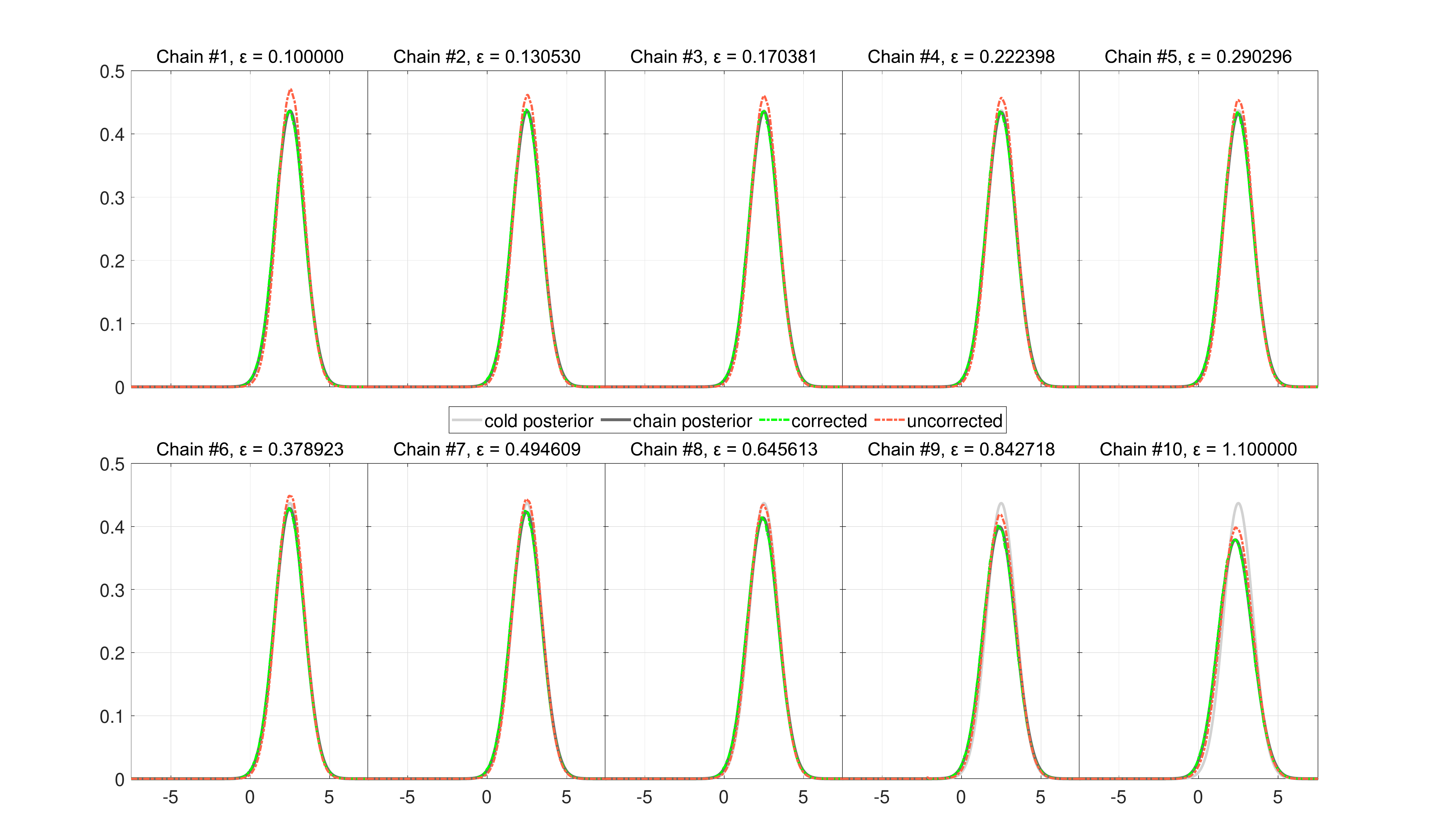}
	\caption[Kernel density estimates of the chains for biased and unbiased runs of the \texttt{ABC-APTMC} algorithm]{Kernel density estimates of all chains for corrected and uncorrected runs of the single processor \texttt{ABC-APTMC} algorithm. In each subplot, the \textit{light grey} line is fixed and represents the cold posterior for reference, the \textit{dark grey} line represents each chain's target posterior (obtained by numerical integration), the dot-dashed \textit{green} lines are kernel density estimates of the chain's posterior returned by the corrected algorithm and are indistinguishable from the dark grey line. The \textit{orange} lines are kernel density estimates for the uncorrected algorithm, and do not agree with the dark grey line, as expected.}
	\label{fig:unormal:posteriors:all}
\end{figure*}
When performing local moves (Algorithm \ref{alg:ABC:1hit}), use a Gaussian random walk proposal with standard deviation $\xi=0.5$. The real-time Markov jump process is run using $\Lambda = 10$ chains. The algorithm is run on a single processor for one hour or $T=3600$ seconds in real time after a $30$ second burn-in, with exchange moves occurring every $\delta_T=5 \times 10^{-4}$ seconds (or $0.5$ milliseconds). The radii of the balls $\varepsilon^{1:\Lambda}$ are defined to vary between  $\varepsilon^1=0.1$ and $\varepsilon^{\Lambda}=1.1$.\par 

We verify that bias correction must be applied for all chains to converge to the correct posterior. This is done by visually comparing density estimates of each of the post burn-in chains to the true corresponding posterior (obtained by numerical integration). When bias correction is not applied, the \texttt{ABC-APTMC} algorithm does not exclude the currently working chain $j$ in its exchange moves. In this case, every chain converges to an erroneous distribution which overestimates the mode of its corresponding posterior, as is visible in Figure \ref{fig:unormal:posteriors:all}. On the other hand, correcting the algorithm for such bias ensures that every chain converges to the correct corresponding posterior. \par
Next, we compare the performance of the \texttt{ABC-APTMC} algorithm to that of a standard ABC (referred to as standard \texttt{ABC}) algorithm. For that, a more applied parameter estimation example is considered, for which the adoption of a likelihood-free approach is necessary.

\subsection[Stochastic Lotka-Volterra model]{Stochastic Lotka-Volterra model} \label{sec:ABC:LV}
In this section, we consider the stochastic Lotka-Volterra predator-prey model (\cite{lotka1926elements}, \cite{volterra1927variazioni}), a model which has been explored in the past (\cite{wilkinson2011stochastic, boys2008bayesian}), including in an ABC setting (\cite{lee2014variance, fearnhead2012constructing, toni2009approximate, prangle2017adapting}).  Let $X_{1:2}(t)$ be a bivariate, integer-valued pure jump Markov process with initial values $X_{1:2}(0) = (50, 100)$, where $X_1(t)$ represents the number of prey and $X_2(t)$ the number of predators at time $t$. For small time interval $\Delta t$, we describe the predator-prey dynamics in the following way:
\begin{align*}
	\mathbb{P}\left\{X_{1:2}(t+\Delta t) = z_{1:2} \,|\, X_{1:2}(t) = x_{1:2}\right\} 
	&= \begin{cases} \theta_1 x_1 \Delta t + o(\Delta t), & \mbox{if } z_{1:2} = (x_1+1, x_2),\\
		\theta_2 x_1 x_2 \Delta t + o(\Delta t), & \mbox{if } z_{1:2} = (x_1-1, x_2+1), \\
		\theta_3 x_2 \Delta t + o(\Delta t), & \mbox{if } z_{1:2} = (x_1, x_2-1), \\
		o(\Delta t), & \mbox{otherwise}.\end{cases}
\end{align*}
In this example, the only observations available are the number of prey, i.e. $X_1$ at 10 discrete time points. Following theory in \cite{wilkinson2011stochastic} (Chapter 6), the process can be simulated and discretised using the \cite{gillespie1977exact} algorithm, in which the inter-jump times follow an exponential distribution. The observations employed were simulated in \cite{lee2014variance} with true parameters $\theta = (1, 0.005, 0.6)$, giving $y = \left\{88, 165, 274, 268, 114, 46, 32, 36, 53, 92\right\}$ at times $\left\{1, \ldots, 10\right\}$.\par This case study presents various challenges. The first challenge is that the posterior is intractable, and some of the components of the parameters $\theta:=\theta_{1:3}$ (namely $\theta_2$ and $\theta_3$) exhibit strong correlations. Therefore, ABC must be employed, and the `ball' considered takes the following form for $\varepsilon>0$:
\begin{align}
	B_{\varepsilon}(y) \nonumber 
	&= \left\{X_1(t) : \left|\log\left[X_1(i)\right]-\log\left[y(i)\right]\right|\leq \varepsilon, \forall i =1, \ldots, 10\right\}.
\end{align}
So a set of simulated $X_1(t)$ is considered as `hitting the ball' if all 10 simulated data points are at most $e^\varepsilon$ times (and at least $e^{-\varepsilon}$ times) the magnitude of the corresponding observation in $y$. \par


In \cite{lee2014variance} (Algorithm 3), the 1-hit MCMC kernel (referred to here as standard \texttt{ABC}) is shown to return the most reliable results by comparison with other MCMC kernels, which are not considered here. The second challenge is that while this algorithm can be reasonably fast, it is highly inefficient as it has a very low acceptance rate, and thus the autocorrelation between samples for low lags is very high. \par 
We have established that the race step in Algorithm \ref{alg:ABC:1hit} takes a random time to complete. In addition to that, its hold time distribution for the Lotka-Volterra model is a mixture between quick and lengthy completion times, as the simulation steps within the 1-hit kernel race are capable of taking a considerable amount of time despite often being almost instant. Indeed, when simulating observations using the discretised Gillespie algorithm, if the number of predators is low, prey numbers will flourish and the simulation will take longer. Hence, the third challenge in this particular model is that the race can sometimes get stuck for extended periods of time if the number of prey to simulate is especially high. Therefore, we aim to first of all improve performances by introducing exchange moves on a single processor (\texttt{ABC-PTMC}). Then $-$ and most importantly $-$ we further improve the algorithm by implementing it within the Anytime framework (\texttt{ABC-APTMC}), a method which works especially well on multiple processors.

\subsubsection{One processor}

The first part of this case study is run on a single processor and serves to demonstrate the gain in efficiency introduced by the exchange moves described in Algorithm \ref{alg:ABC:exch:fast}. Define the prior on $\theta \in \left[0, \infty\right)^3$ for the single processor experiment to be 
$p(\theta) = \exp\left\{-\theta_1 - \theta_2 - \theta_3\right\}$. The proposal distribution is a truncated normal, i.e. $\theta' \, | \, \theta \sim \mathcal{TN}(\theta, \Sigma)$, $\theta' \in (0, 10)$ with mean $\theta$ and covariance $\Sigma =\text{diag}\left(0.25, 0.0025, 0.25\right)$.
The truncated normal is used in order to ensure that all proposals remain non-negative. For reference, $2364$ independent samples from the posterior are obtained via ABC rejection sampling with $\varepsilon=1$ and the density estimates in Figure 6 of \cite{lee2014variance} are reproduced. To obtain these posterior samples, $10^7$ independent samples from the prior were required, yielding the very low $0.024\%$ acceptance rate. This method of sampling from the posterior is therefore extremely inefficient, and the decision to resort to MCMC kernels in order to improve efficiency is justified.  \par

On a single processor, the three algorithms considered are the vanilla 1-hit MCMC kernel (standard \texttt{ABC}) defined in Algorithm \ref{alg:ABC:1hit}, the single processor version of the algorithm with added exchange moves (\texttt{ABC-PTMC-1}) and the same but within the Anytime framework (\texttt{ABC-APTMC-1}). They are run nine times for 100800 seconds (28 hours)  $-$ after 3600 seconds (1 hour) of burn-in $-$ and their main settings are summarised in Table \ref{table:LV:settings:single}. \par 
Given the aim is to compare the performance of these algorithms, it is important to note that the parallel tempering algorithms, having to deal with updating multiple chains sequentially, are likely to return cold chains with fewer samples. The algorithms must therefore be properly set up such that the gain in efficiency introduced by exchange moves is not overshadowed by the greater number of chains and computational cost of having to update them all. In this experiment, the parallel tempering algorithms are run on $\Lambda=6$ chains, each targeting posteriors associated with balls of radii $\varepsilon^{1:6} = \left\{1, 1.1447, 1.3104, 1.5, 11, 15\right\}$ and the proposal distribution has covariance $\Sigma^{1:6}$ where $\Sigma^{\lambda} = \text{diag}\left(\sigma^{\lambda} , \sigma^{\lambda}  10^{-2}, \sigma^{\lambda} \right)$ and $\sigma^{1:6} = \left\{0.008, 0.025, 0.05, 0.09, 0.25, 0.5\right\}$. Exchange moves are performed as described in Algorithm \ref{alg:ABC:exch:fast} and 
alternate between odd $(1, 2), (3, 4), (5, 6)$ (excluding $(5, 6)$ in the Anytime version) and even $(2, 3), (4, 5)$ pairs of eligible chains. 
As per Section \ref{sec:APTMC:tuning}, exchange moves for the \texttt{ABC-PTMC-1} algorithm are performed every $\Lambda=6$ local moves, and the real-time deadline $\delta$ for exchange moves in the \texttt{ABC-APTMC-1} algorithm is determined by repeatedly measuring the times taken by the \texttt{ABC-PTMC-1} algorithm to perform these $6$ moves and setting $\delta$ to be the median over all measured times. \par

\begin{table*}[htbp]                                          
	\centering           
	{\def\arraystretch{1}
		\begin{tabular}{c c c c c c}  
			\toprule
			\textit{\textbf{Label}}		& \textit{\textbf{Workers}}	& \textit{\textbf{Chains}} & \textit{\textbf{Chains per}}	& \textit{\textbf{Exchange moves}} & \textit{\textbf{Anytime}}\\ 
			& $W$	& $\Lambda$ & \textit{\textbf{worker}} $K$	& \textit{\textbf{(every)}}  & \\ 
			\midrule
			\texttt{ABC}							&	1 	& 1	& 1		& none 									& No	\\	
			\texttt{ABC-PTMC-1}				&	1		& 6	& 6		& 6 local moves		& No	\\ 	
			\texttt{ABC-APTMC-1}			&	1		& 6 & 6		& $2.59$ seconds	& Yes	\\			
			\bottomrule                                                  
		\end{tabular}  
	}
	\caption{Algorithm information and settings for stochastic Lotka-Volterra predator-prey model on a single processor.}
	\label{table:LV:settings:single}                              
\end{table*} 

\subsubsection{Multiple processors}
\label{settings:LV:multi}
Next, we demonstrate the gain in efficiency introduced by running the parallel tempering algorithm within the Anytime framework on multiple processors. 
The algorithms considered are the multi-processor \texttt{ABC-PTMC-W} and \texttt{ABC-APTMC-W} algorithms and their single processor counterparts \texttt{ABC-PTMC-1} and \texttt{ABC-APTMC-1} . This time, we define a uniform prior between 0 and 3. The proposal distribution is still a truncated normal, but with tighter limits (corresponding to the prior) i.e. $\theta' \, | \, \theta \sim T\mathcal{N}(\theta, \Sigma)$, $\theta' \in (0, 3)$. \par 
The two algorithms are run on $\Lambda = 20$ chains, each targeting posteriors associated with balls of radii ranging from $\varepsilon^{1} = 1$ to $\varepsilon^{20} = 11$ and proposal distribution covariances $\Sigma^{1:20}$ where $\Sigma^{\lambda} = \text{diag}\left(\sigma^{\lambda} , \sigma^{\lambda}  10^{-2}, \sigma^{\lambda} \right)$ for chain $\lambda=1, \ldots, 20$ and where values range from $\sigma^1 =  0.008$ to $\sigma^{20} = 0.5$ (see Table \ref{table:LV:multi:samplesizes}). 
The algorithms are run four times for 864000 seconds (24 hours) and their main settings are summarised in Table \ref{table:LV:settings:multi}. Given the non-negligible 1.1 second communication overhead, this experiment is run according to the third scenario from Section \ref{sec:APTMC:implementation}, i.e. dividing exchange moves into within- and between-worker exchange moves. As described in Section \ref{sec:APTMC:tuning}, a full set of parallel moves here consists of $K=5$ local moves, followed by within-worker exchange moves between a pair of adjacent chains selected at random, followed by $5$ more local moves. The between-worker exchange moves are performed after a full set of parallel moves on the master by selecting a pair of adjacent workers at random and exchanging between the warmest eligible chain from the first worker and coldest from the second so that they are adjacent. \par	
\begin{table*}[htbp]                                          
	\centering           
	\resizebox{\columnwidth}{!}{\begin{tabular}{ccccclc}  
			\toprule
			\textit{\textbf{Label}}		& \textit{\textbf{Workers}}	& \textit{\textbf{Chains}} & \textit{\textbf{Chains per}}	& \textit{\textbf{Communication}} & \textit{\textbf{Exchange moves (every)}} & \textit{\textbf{Anytime}}\\ 
			& $W$	& $\Lambda$ & \textit{\textbf{worker}} $K$ & \textit{\textbf{overhead}}	&  & \\ 
			\midrule
			\texttt{ABC-PTMC-1}	 &	1	& 20 & 20 & -		& $20$ local moves	& No	\\ 	
			\texttt{ABC-APTMC-1} &	1	& 20 & 20 & -		& $11$ seconds	& Yes\\
			\texttt{ABC-PTMC-W}	 &	4	& 20 & 5  & 1.1 seconds & $5$ local moves	& No	\\ 	
			\texttt{ABC-APTMC-W} &	4	& 20 & 5  & 1.1	seconds	& $5$ local moves (\textit{within} workers)	& Yes\\
			&		&  &   &	& $15.3$ seconds (\textit{between} workers)	& \\
			\bottomrule                                                  
	\end{tabular}  }
	\caption{Algorithm information and settings for stochastic Lotka-Volterra predator-prey model on multiple processors.}
	\label{table:LV:settings:multi}                              
\end{table*} 

\subsubsection{Performance evaluation}\label{results:LV:all}

All algorithms returned density estimates that were close to those obtained via rejection sampling. In order to compare the performance of the algorithms, they are set to run for the same real time period. The $IAT$ and cumulative $ESS$ over all repeat runs are computed for all algorithms.
The $ESS$ is particularly important here, as it gives us how many effective samples the different algorithms can return within a fixed time frame. For example, a very fast algorithm could still return a higher $ESS$ even if it has a much higher $IAT$. Additionally, to illustrate how the Anytime version of the parallel tempering algorithms works compared to standard \texttt{ABC-PTMC}, the real times all algorithms take to perform local and exchange moves are measured and their timelines plotted in Figure \ref{fig:LV:both}. \par

\begin{figure*}[htbp]
	\centering
	\includegraphics[width=\textwidth]{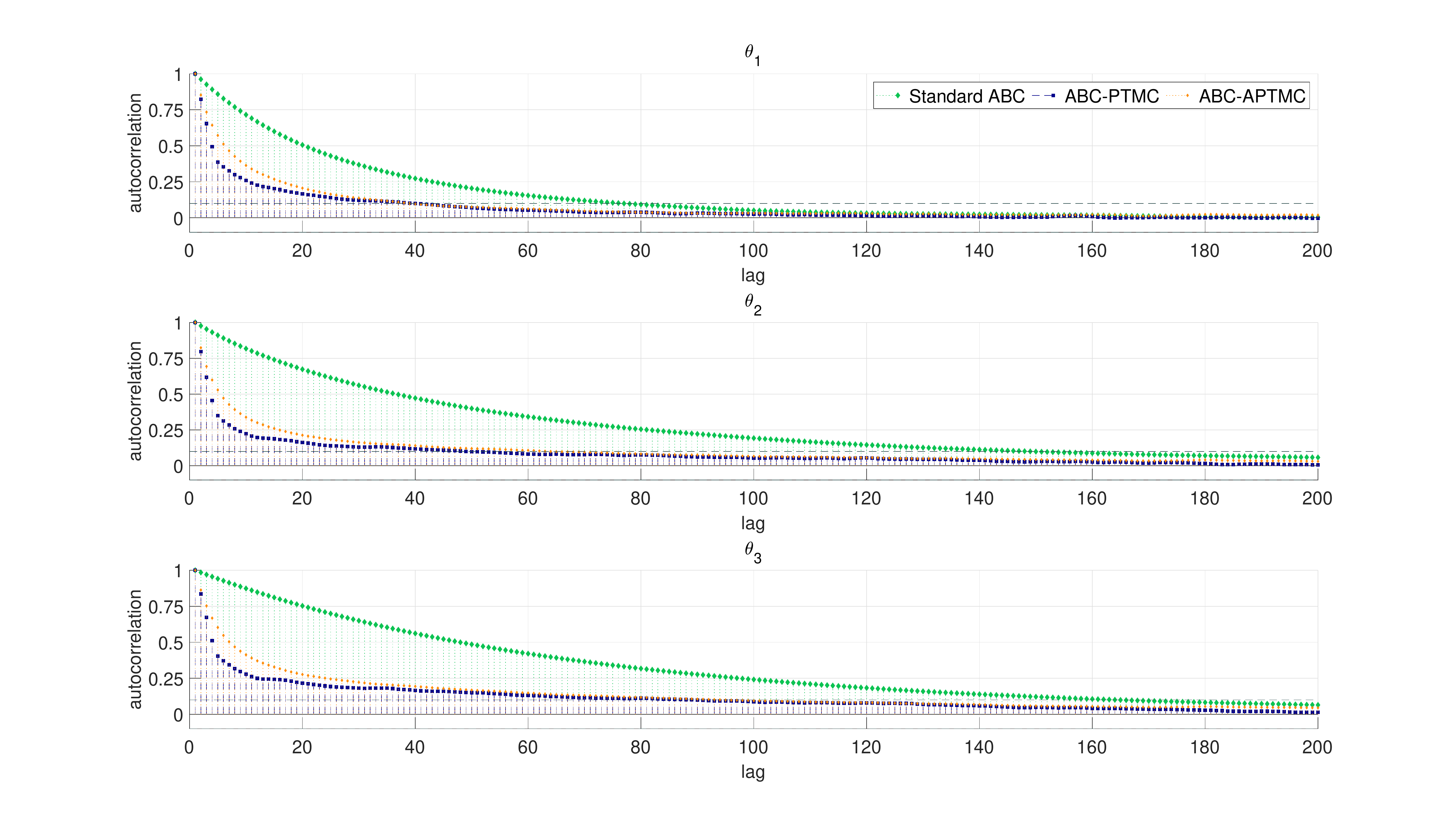}
	\caption[Mean Sample acf of the cold chain for nine runs of the standard \texttt{ABC}, \texttt{ABC-PTMC-1} and \texttt{ABC-APTMC-1} algorithms]{Plots of the sample autocorrelation function up to lag 200 of the cold chain for runs of the standard \texttt{ABC} (\textit{green}), \texttt{ABC-PTMC-1} (\textit{blue}) and \texttt{ABC-APTMC-1} (\textit{orange}) algorithms to estimate the posterior distributions of the parameters $\theta = \left(\theta_1, \theta_2, \theta_3\right)$ of a stochastic Lotka-Volterra model. The inclusion of exchange moves boosts efficiency and leads to a steeper decay in the parallel tempering algorithms.}
	\label{fig:LV:acf:1}
\end{figure*}

\begin{figure*}[htbp]
	\centering
	\includegraphics[width=\textwidth]{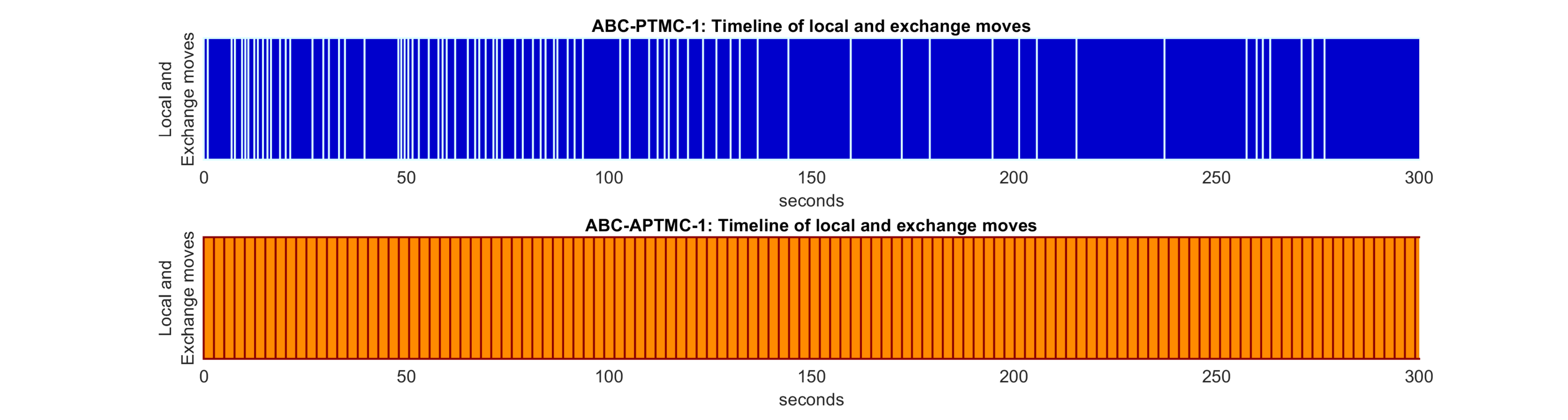}
	\caption[Timeline of local and exchange moves for the \texttt{ABC-PTMC-1} algorithm]{Timeline of local and exchange moves for the \texttt{ABC-PTMC-1} and \texttt{ABC-APTMC-1} algorithms for the first 300 seconds. The exchange moves are represented by the \textit{white} and \textit{red} lines and the local moves by the \textit{dark blue} and \textit{orange} coloured blocks. Local moves take a random amount of time to complete, as illustrated by the times consumed by local moves for the  \texttt{ABC-PTMC-1} algorithm. The Anytime (\texttt{ABC-APTMC-1}) version effectively implements a hard deadline for the exchange moves (without introducing a bias), as can be seen by the regularity of local move times in the bottom figure.} 
	\label{fig:LV:both}
\end{figure*}

\paragraph{One processor}
\label{results:LV:one}	
Both the \texttt{ABC-PTMC-1} and \texttt{ABC-APTMC-1} algorithm display an improvement in performances: they return $IAT$s that are respectively 3.2 and 1.6 times lower on average than those of the standard \texttt{ABC} algorithm in Table \ref{table:LV:acf:1}, and display a steeper decay in sample autocorrelation in Figure \ref{fig:LV:acf:1}. In the 28 hours (post burn-in) during which the algorithms ran, both parallel algorithms also yielded an increased $ESS$. The effect of the Anytime framework on the behaviour of the parallel tempering algorithm is demonstrated in Figure \ref{fig:LV:both}. The timeline of local moves for the \texttt{ABC-PTMC-1} algorithm illustrates the fact that local moves take a random amount of time to complete. In the Anytime version of the algorithm, this is mitigated since a deadline was implemented. As a result, the bottom plot in Figure \ref{fig:LV:both} displays more consistent local move times. \par
Note that in Table \ref{table:LV:acf:1}, while the improvement in $IAT$ is significant, the increase in $ESS$ after 28 hours is not particularly huge. This is due to the previously mentioned erratic behaviour of the hold time distribution for this example. Other examples explored such as the moving average example in \cite{marin2012approximate} (not reported here) yielded a much more significant increase in $ESS$ after introducing exchange moves. We also note that in this example, the \texttt{ABC-PTMC-1} algorithm returned a lower $IAT$ than its Anytime counterpart but Anytime had a larger $ESS$. There are two potential reasons to account for the $IAT$. The first is the many swaps which are cycling the same samples among the held chains of Anytime. The second, as mentioned in Section \ref{sec:APTMC:tuning}, is that the Anytime algorithms cannot always exchange the samples of adjacent chains, because they must exclude the chain that is currently computing, and this causes a slightly higher rejection rate compared to the standard version (in the multi-processor example with more chains, this is mitigated). However, the many more swap moves of Anytime does then result in more returned samples, which leads to a higher $ESS$. The single processor experiment was mainly designed to demonstrate the performance improvements brought by adding exchange moves to the 1-hit MCMC kernel (referred to as standard \texttt{ABC}) and to show that Anytime does match the performance of parallel tempering on a single processor. Since a single processor is never idling, the strength of the Anytime framework is best illustrated in a multi-processor setting. \par

\begin{figure*}[htbp]
	\centering
	\includegraphics[width=\textwidth]{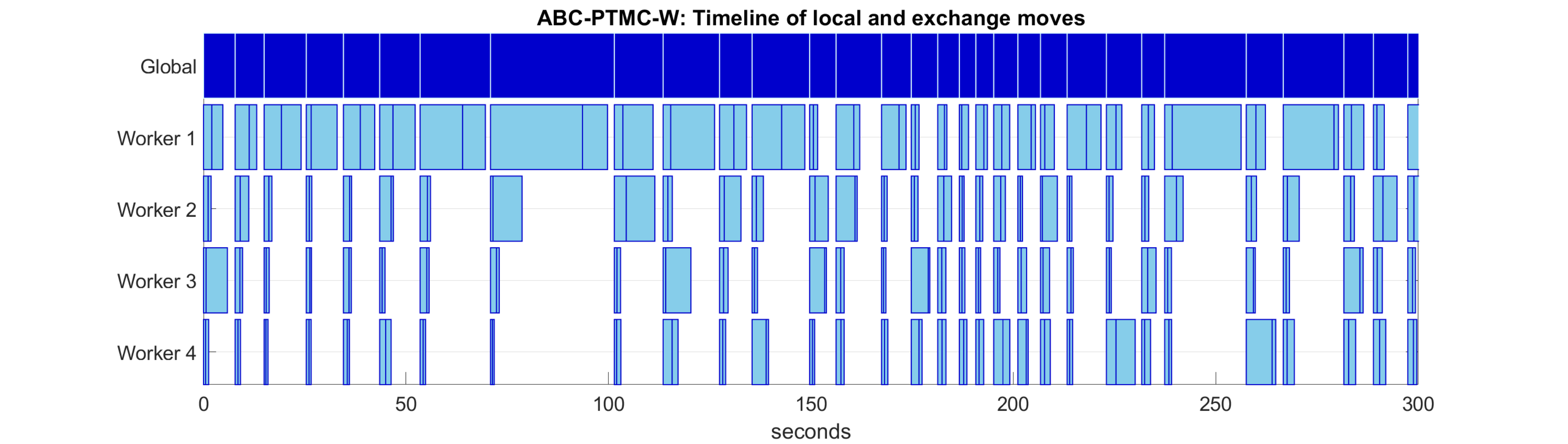}
	\caption[Timeline of local and exchange moves for the \texttt{ABC-PTMC-W} algorithm]{Timeline of local and exchange moves for the \texttt{ABC-PTMC-W} algorithm  for the first 300 seconds. Within and between worker exchange moves are represented by the \textit{white lines} on the Global timeline and \textit{blue lines} on the various Worker timelines, respectively. Local moves on each worker are represented by the \textit{light blue} coloured blocks and the \textit{dark  blue} coloured blocks correspond to the global time all workers spend running in parallel, including communication overhead. Significant idle time is apparent on all workers as they always have to wait for the slowest among them to complete its set of local moves.}
	\label{fig:LV:TM:K}
\end{figure*}

\begin{table*}[htbp]                                          
	\centering           
	{\def\arraystretch{1.3}
		\begin{tabular}{c cc cc cc}  
			\toprule
			& \multicolumn{2}{c}{Standard \texttt{ABC}} & \multicolumn{2}{c}{\texttt{ABC-PTMC-1}} & \multicolumn{2}{c}{\texttt{ABC-APTMC-1}} \\              
			
			& $IAT$ 	& $ESS$ 	& $IAT$		& $ESS$ 	& $IAT$ 	& $ESS$ \\ 																							
			\midrule                                                 
			$\theta_1$	& 69.476 	& 7018.1	& 22.404	& 7618.6	& 44.071	& 7963.8\\ 
			$\theta_2$	& 122.73	& 3973		& 35.381	& 4824.2	& 69.803 	& 5028\\
			$\theta_3$	& 150.74	& 3234.6	& 50.035	& 3411.3	& 98.929	& 3547.7\\ 
			\bottomrule                                                  
		\end{tabular}  
	}
	\caption[$ESS$ and $IAT$ over nine 28-hour runs of the \texttt{ABC}, \texttt{ABC-PTMC-1} and \texttt{ABC-APTMC-1} algorithms.]{Integrated autocorrelation time ($IAT$) and cumulative effective sample size ($ESS$) over nine 28-hour runs of the standard \texttt{ABC}, \texttt{ABC-PTMC-1} and \texttt{ABC-APTMC-1} algorithms to estimate the posterior distributions of the parameters $\theta = \left(\theta_1, \theta_2, \theta_3\right)$ of a stochastic Lotka-Volterra model. Improvements in performance are modest in this example.}
	\label{table:LV:acf:1}                              
\end{table*} 
\paragraph{Multiple processors}
\label{results:LV:multi}	
In the multi-processor case study, both the \texttt{ABC-PTMC-1} and \texttt{ABC-PTMC-W} were set so that on each worker, an exchange move occurred after all chains had been updated locally once, as described in Table \ref{table:LV:settings:multi}. The total number of samples returned by the \texttt{ABC-PTMC-W} algorithm is higher for all chains (see Table \ref{table:LV:multi:samplesizes}).
However, the \texttt{ABC-PTMC-W} algorithm is just as affected by the distribution of the hold times being a mixture of quick and lengthy completion times as its single processor counterpart, and is just as prone to getting stuck in a race for an extended period. During this time, all processors sit idle while waiting for the race to complete, as illustrated in Figure \ref{fig:LV:TM:K}. Therefore, the \texttt{ABC-PTMC-W} algorithm struggles to properly boost the total sample size output by the cold chain, and the $ESS$ is not markedly higher on average in Table \ref{table:LV:acf:K}. On the other hand, thanks to the real time deadlines implemented, the \texttt{ABC-APTMC-W} algorithm is able to more than double the total size of the samples output (see Table \ref{table:LV:multi:samplesizes}), and the $ESS$s for the cold chain returned in Table \ref{table:LV:acf:K} are on average 3.41 times higher than those of the single processor version. \par
\begin{figure*}[htbp]
	\centering
	\includegraphics[width=\textwidth]{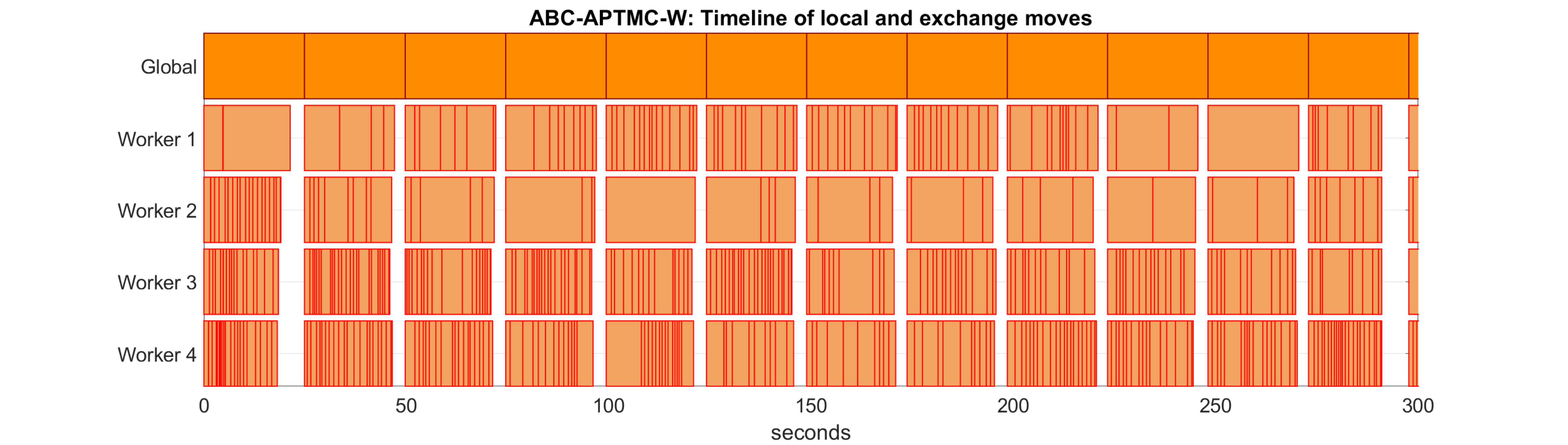}
	\caption[Timeline of local and exchange moves for the \texttt{ABC-APTMC-W} algorithm]{Timeline of local and exchange moves for the \texttt{ABC-APTMC-W} algorithm  for the first 300 seconds. Within and between worker exchange moves are represented by the \textit{red lines}. Local moves on each worker are represented by the various \textit{orange} coloured blocks, with the brighter blocks corresponding to the global time all workers spend running in parallel, including communication overhead. The significant idle time from Figure \ref{fig:LV:TM:K} has been greatly reduced thanks to the deadlines implemented.}
	\label{fig:LV:TM:a:K}
\end{figure*}

As for the main comparison $-$ namely Anytime vs standard ABC with exchange moves $-$ both single and multi-processor Anytime algorithms return an $ESS$ larger than their respective standard versions in Table \ref{table:LV:acf:K}. While the improvement on a single processor is modest, the $ESS$ has more than quadrupled on multiple processors. Figures \ref{fig:LV:TM:K} and \ref{fig:LV:TM:a:K} illustrate well the advantage of implementing a real-time deadline to local moves. At most local moves, the issue in which all workers sit idle waiting for the slowest to finish arises for the \texttt{ABC-PTMC-W} algorithm. On the other hand, Figure \ref{fig:LV:TM:a:K} shows that the Anytime version of the algorithm is making better use of the allocated computational resources. The Anytime framework ensures that none of the workers need to wait for the slowest among them to finish, allowing for more exploration of the sample space in the faster workers. Additionally, the real time deadline ensures that even if chain $k$ on worker $w$ remains stuck in a race for an extended period of time, the other workers are still updating. Therefore, while the remaining four chains on worker $w$ wait for chain $k$ to complete its race, they also continue to be updated at regular intervals thanks to the exchange moves with other workers. The addition of ABC exchange moves in his case study proved fruitful, as the $ESS$ for the parameters of the Lotka-Volterra model was increased. 

\begin{table*}[htbp]                                          
	\centering           
	{\def\arraystretch{1.3}
		\begin{tabular}{c cc cc cc cc}  
			\toprule		& \multicolumn{4}{c}{\textbf{One processor}} & \multicolumn{4}{c}{\textbf{Multiple processors}} \\
			& \multicolumn{2}{c}{\texttt{ABC-PTMC-1}} & \multicolumn{2}{c}{\texttt{ABC-APTMC-1}} & \multicolumn{2}{c}{\texttt{ABC-PTMC-W}} & \multicolumn{2}{c}{\texttt{ABC-APTMC-W}}\\                
			& $IAT$ 	& $ESS$ 	& $IAT$		& $ESS$ 	& $IAT$ 	& $ESS$ 	& $IAT$ 	& $ESS$  \\ 																							
			\midrule                                                 
			$\theta_1$	& 39.535  & 269.89  & 72.475  & 362.62  & 48.621 & 266.7  & 39.898 & 1452.5 \\
			$\theta_2$	& 72.908  & 146.35  & 88.446  & 297.14  & 67.395 & 192.4  & 72.79  & 796.14 \\
			$\theta_3$	& 82.464  & 129.39  & 138.56  & 189.68  & 87.635 & 147.97 & 101.57 & 570.57 \\
			\bottomrule                                                  
		\end{tabular}  
	}
	\caption[$ESS$ and $IAT$ over four 24-hour runs of the \texttt{ABC-PTMC-1}, \texttt{ABC-APTMC-1}, \texttt{ABC-PTMC-W} and \texttt{ABC-APTMC-W} algorithms.]{Integrated autocorrelation time ($IAT$) and cumulative effective sample size ($ESS$) over four 24-hour runs of the \texttt{ABC-PTMC-1}, \texttt{ABC-APTMC-1}, \texttt{ABC-PTMC-W} and \texttt{ABC-APTMC-W} algorithms to estimate the posterior distributions of the parameters $\theta = \left(\theta_1, \theta_2, \theta_3\right)$ of a stochastic Lotka-Volterra model.}
	\label{table:LV:acf:K}                              
\end{table*} 

	\section{Conclusion}
\label{cha:discussion}
In an effort to increase the efficiency of MCMC algorithms, in particular for use on multiple processors, and for situations in which compute times of the algorithms depend on their current states, the APTMC algorithm was developed. The algorithm combines the enhanced exploration of the state space, provided by the between-chain exchange moves in parallel tempering, with control over the real-time budget and robustness to interruptions available within the Anytime Monte Carlo framework. Then, an application of APTMC to problems where the likelihood is unavailable and an ABC MCMC kernel, in particular the 1-hit MCMC kernel, must be employed instead was considered. \par

Initially, the construction of the Anytime Monte Carlo algorithm with the inclusion of exchange moves on a single and multiple processors was verified on a Gamma mixture example. The performance improvements they brought were then demonstrated by comparing the algorithm to a standard MCMC algorithm. Subsequently, the exchange moves were adapted for pairing with the 1-hit MCMC kernel, a simulation-based algorithm within ABC framework, which provides an attractive, likelihood-free approach to MCMC. The construction of the adapted ABC algorithm was verified using a  univariate normal example. Then, the increased efficiency of the inclusion of exchange moves was demonstrated in comparison to that of a standard ABC algorithm on a parameter estimation problem. The problem involved the parameters of a stochastic Lotka-Volterra predator-prey model based on partial and discrete data, and the likelihood of this model is intractable. On a single processor, it was shown that introducing exchange moves provides an improvement in performance and an increase in the effective sample size compared to that of the standard, single chain ABC algorithm. The Anytime results for a single processor matches the efficiency of standard parallel tempering, which is to be expected since the single processor is never idling in both the Anytime and non-Anytime versions. The $ESS$ gains of Anytime become significant in a multi-processor setting since one slow processor will lead to all the others idling in standard parallel tempering. \par
One major class of applications with local moves that have state-dependent real completion times and could benefit from the APTMC framework are transdimensional applications, such as RJ-MCMC (\cite{green1995reversible}), which has a parallel tempering implementation in \cite{jasra2007population}. The performance of parallel tempering algorithms with temperature-dependent completion times, as addressed in \cite{earl2004optimal}, can also be improved by the Anytime framework. Examples of such algorithms occur in \cite{hritz2007optimization, karimi2011high, wang2003parallel}. From a purely computing infrastructure-related perspective, exogenous factors such as processor hardware, competing jobs, memory bandwidth, network traffic or I/O load also affect the completion time of local moves even if they are not state-dependent within the algorithm itself. This is the case in \cite{rodinger2006distributed}. In a more general setting, any population-based MCMC algorithm such as parallel tempering, SMC samplers (\cite{del2006sequential}), or parallelised generalised elliptical slice sampling (\cite{nishihara2014parallel}), which combines a parallel updates step (e.g. local moves)  and an inter-processor communication step (e.g. exchange moves, resampling) can benefit from the APTMC framework in future implementations.\par
As a final comment, we note the potential relevance of the work of  \cite{dupuis2012infinite} in studying efficiency as a function of exchange frequency. As exchange steps of  Anytime parallel tempering become more frequent, i.e. many occur between the stalled chains before the local move completes, it would be interesting to explore if our Anytime parallel tempering algorithm could be understood in the framework of the \emph{infinite} swapping limit version of parallel tempering which has been shown in  \cite{dupuis2012infinite} to dominate in numerical examples and in a specific theoretical context. However, their analysis ignores the cost of performing exchanges, which is non-negligible when communicating across processors, and thus cannot be plainly advocated without more consideration.

\begin{table*}[htbp!]                                                                     
	\centering                                                                        
	\begin{tabular}{c c c c c c c}                                                  
		\toprule
		Chain $k$ & $\varepsilon^k$ & $\sigma^k$ & \texttt{ABC-PTMC-1} & \texttt{ABC-PTMC-W} & \texttt{ABC-APTMC-1} & \texttt{ABC-APTMC-W} \\		
		\midrule                                                                          
		1   & 1      & 0.008  &       2667.5           &       3241.8           &       6570.3           &        14488           \\
		2   & 1.046  & 0.009  &         2790           &       3564.8           &       6934.8           &        16902           \\
		3   & 1.094  & 0.011           &       2796.8           &       3567.5           &         6941           &        16837           \\
		4   & 1.145  & 0.012           &       2797.3           &       3604.5           &       6924.8           &        17264           \\
		5   & 1.197  & 0.014           &       2793.8           &       3719.8           &       6931.3           &        15710           \\
		\hline
		6   & 1.253  & 0.016           &       2786.3           &       3748.8           &       6951.5           &        17157           \\
		7   & 1.31   & 0.019           &       2784.8           &       3629.3           &       6961.3           &        18947           \\
		8   & 1.371  & 0.022           &       2795.5           &         3615           &       6941.5           &        18551           \\
		9   & 1.434  & 0.025           &       2805.5           &       3608.5           &       6950.8           &        18759           \\
		10  & 1.5    & 0.029           &       2803.5           &       3711.5           &       6962.8           &        17276           \\
		\hline
		11  & 1.661  & 0.034           &       2798.5           &       3799.8           &       6962.3           &        46350           \\
		12  & 1.84   & 0.039           &       2803.3           &       3693.3           &         6983           &        53716           \\
		13  & 2.038  &        0.045           &       2814.8           &       3656.5           &       6995.3           &        53289           \\
		14  & 2.257  &        0.052           &         2799           &       3658.3           &       7008.8           &        53458           \\
		15  & 2.5    &         0.06           &       2783.5           &       3796.3           &       7029.8           &        46597           \\
		\hline
		16  & 3.362  &        0.092           &       2787.5           &       4054.5           &       7038.5           &        68953           \\
		17  & 4.522  &         0.14           &       2783.8           &       3936.5           &       7009.5           &        79231           \\
		18  & 6.082  &        0.214           &         2781           &       3912.5           &       6982.5           &        78917           \\
		19  & 8.179  &        0.327           &       2780.8           &       3919.5           &       7002.8           &        79038           \\
		20  & 11     &          0.5           &       2665.8           &       3598.5           &       6604.3           &        67725           \\ 
		\bottomrule                                                                            
	\end{tabular}                                                                     
	\caption[Average sample sizes per chain returned over four 24-hour runs of the \texttt{ABC-PTMC-1}, \texttt{ABC-APTMC-1}, \texttt{ABC-PTMC-W}, \texttt{ABC-APTMC-W} algorithms.]{Average sample sizes per chain returned over four 24-hour runs of the \texttt{ABC-PTMC-1}, \texttt{ABC-APTMC-1}, \texttt{ABC-PTMC-W}, \texttt{ABC-APTMC-W} algorithms to estimate the posterior distributions of the parameters $\theta$ of a stochastic Lotka-Volterra model on multiple processors in Section \ref{results:LV:all}. The ball radius $\varepsilon^k$ and proposal distribution covariance diag$\left(\sigma^k, \sigma^k 10^{-2}, \sigma^k\right)$ associated with each chain $k$ are displayed for information.}
	\label{table:LV:multi:samplesizes}                                                        
\end{table*}         
\newpage

%
%

\bibliographystyle{abbrvnat}
\bibliography{BibTeX1}   

%
%
\newpage
\appendix
\section{Proofs}
	
	\subsection{Proof of Proposition  \ref{prop:anytime:exch}} \label{app:proof:prop}
	The continuous time chain of Proposition \ref{prop:anytime:exch} is described in steps \ref{alg:APTMC:one:1} to \ref{alg:APTMC:one:6} (excluding the exchange steps) of Algorithm \ref{alg:APTMC:one}. The Markov transition
	kernel of this chain is		
	\begin{alignat*}{1}
		&(P_{t}f)(x^{1:\text{\ensuremath{\Lambda}}},l,j) \\
		&=\mathbb{E}\left\{ f(X^{1:\Lambda}(t),L(t),J(t))\vert\left(X^{1:\Lambda},L,J\right)(0)=\left(x^{1:\text{\ensuremath{\Lambda}}},l,j\right)\right\} \\
		& =\mathbb{E}\left\{ f(X^{1:\Lambda}(t),L(t),J(t))\mathbb{I}_{\left\{ L(t)\geq t\right\} }\vert x^{1:\text{\ensuremath{\Lambda}}},l,j\right\} 
		+\mathbb{E}\left\{ f(X^{1:\Lambda}(t),L(t),J(t))\mathbb{I}_{\left\{ L(t)<t\right\} }\vert x^{1:\text{\ensuremath{\Lambda}}},l,j\right\}, 
	\end{alignat*}
	where in the second line, the conditioning on the state at time 0 has	been abbreviated, and the two events $\left\{ L(t)\geq t\right\} $ and $\left\{ L(t)<t\right\} $ have been introduced to simplify the calculation. \\
	The event $\left\{ L(t)\geq t\right\} $ implies that chain $j$ hasn't completed its local move by time $t$. It follows then that 
	\begin{alignat*}{1}
		 \mathbb{E}\left\{ f(X^{1:\Lambda}(t),L(t),J(t))\mathbb{I}_{\left\{ L(t)\geq t\right\} }\vert x^{1:\text{\ensuremath{\Lambda}}},l,j\right\}  
		& =f(x^{1:\Lambda},l+t,j)\frac{\bar{F}_{j}(l+t\vert x^{j})}{\bar{F}_{j}(l\vert x^{j})},
	\end{alignat*}
	where $\bar{F}_{j}(l\vert x^{j})=1-F_{j}(l\vert x^{j})$ and $F_{j}(l\vert x^{j})$ is the cdf of the hold time density of $\tau_{j}(h\vert x^{j})\mathrm{d}h$ for chain $j$. Note that the conditioning on $(x^{1:\text{\ensuremath{\Lambda}}},l,j)$ 		gives rise to the term $\bar{F}_{j}(l\vert x^{j})$ in the denominator, and thus the ratio is the probability $\mathbb{P}\left(L(t)\geq t\vert x^{1:\text{\ensuremath{\Lambda}}},l,j\right)$.\par		
	To simplify the calculation for the event $\left\{ L(t)<t\right\} $, assume $t\leq\epsilon$ where $\epsilon$ is the assumed (in Section \ref{cha:literature}) minimum hold time. That is, the hold time (variable $L(t)$ here) exceeds $\epsilon>0$ with probability 1. Thus, the event $\left\{ L(t)<t\right\}$ for $t\leq\epsilon$ corresponds to a single possible scenario where chain $j$ completes its local move at some time $s$ in the time
	interval $(0,t]$, and thus holds for a total time of $l+s$. Chain
	$j+1$ is next to be worked on, and is still being worked on at time
	$t$, thus $J(t)=j+1$ and $L(t)=t-s$. Applying this reasoning, we
	have 
	\begin{alignat*}{1}
		& \mathbb{E}\left\{ f(X^{1:\Lambda}(t),L(t),J(t))\mathbb{I}_{\left\{ L(t)<t\right\} }\vert x^{1:\text{\ensuremath{\Lambda}}},l,j\right\} \\
		& =\int_{0}^{t}\left(\int(P_{t-s}f)(x^{1:j-1},y,x^{j+1:\Lambda},0,j+1) \kappa_{j}(y\vert x^{j},l+s)\mathrm{d}y\right) 
		\times \frac{\tau_{j}(l+s\vert x^{j})}{\bar{F}_{j}(l\vert x^{j})}\mathrm{d}s,
	\end{alignat*}
	where the inner integral averages over the new state for chain $x^{j}\rightarrow y$ when the hold time is $l+s$, while other states $x^{1:j-1}$ and $x^{j+1:\Lambda}$ are unchanged. The outer integral averages over the hold time distribution conditioned on $L(0)=l$. The usual semigroup property (see \cite{del2017stochastic} for general background) for a Markov process $\left(P_{t}f\right)(x^{1:\text{\ensuremath{\Lambda}}},l,j)=\left(P_{s}(P_{t-s}f)\right)(x^{1:\text{\ensuremath{\Lambda}}},l,j)$,
	is also being employed. \par
	A final simplification is applied to the integrand
	\begin{alignat*}{1}
		& (P_{t-s}f)(x^{1:j-1},y,x^{j+1:\Lambda},0,j+1) 
		 =f(x^{1:j-1},y,x^{j+1:\Lambda},t-s,j+1)
	\end{alignat*}
	since $t-s\leq\epsilon$ and thus the hold time of chain $j$ advances
	to $t-s$ from $0$. \par
	Using the specific form of $A(\mathrm{d}x^{1:\Lambda},\mathrm{d}l,j)$ given in Proposition \ref{prop:anytime:exch} and the integrand $(P_{t}f)(x^{1:\text{\ensuremath{\Lambda}}},l,j)$
	developed above, gives the desired result, namely for any $t\leq\epsilon$, we have
	\begin{alignat*}{1}
		\sum_{j=1}^{\Lambda}\int(P_{t}f)(x^{1:\text{\ensuremath{\Lambda}}},l,j)A(\mathrm{d}x^{1:\Lambda},\mathrm{d}l,j) 
		 =\sum_{j=1}^{\Lambda}\int f(x^{1:\text{\ensuremath{\Lambda}}},l,j)A(\mathrm{d}x^{1:\Lambda},\mathrm{d}l,j).
	\end{alignat*}
	The results thus generalises to any $t>\epsilon$ by the semigroup
	property: $\left(P_{t+h}f\right)=P_{h}\left(P_{t}f\right)$.

\subsection{Anytime distribution of the cold chain} \label{app:proof:anytime}

	To obtain the anytime distribution in the Gamma mixture example in Section \ref{cha:APTMC:GM}, compute the three components of the expression in Equation (\ref{eq:anytime}):
	
	\begin{enumerate}
		\item The density of $X$
			\begin{equation*}
				\pi(\text{d}x) = \frac{x^{k_1-1}}{2 \Gamma(k_1) \theta_1^{k_1}} e^{-\frac{x}{\theta_1}} 
				+ \frac{x^{k_2-1}}{2 \Gamma(k_2) \theta_2^{k_2}} e^{-\frac{x}{\theta_2}} \, \text{d}x,
			\end{equation*}
			where $\Gamma(\,\cdot\,)$ is the gamma function.
		\item The expectation of $H \, | \, x$ given by 
		\begin{equation*}
		\mathbb{E}\left[H \, | \, x\right] = \psi x^p + (1-\psi)x^p = x^p.
		\end{equation*}		
			The $\psi$ factors cancel out, meaning that the anytime distribution is independent of $\psi$ and therefore its value can be chosen to be $1$ for convenience.
		\item To compute $\mathbb{E}\left[H\right]$, use a property of conditional expectation and the honesty conditions of the Gamma$\left(k_1+p, \theta_1\right)$ and Gamma$\left(k_2+p, \theta_2\right)$ distributions:
			\begin{align*}
				\mathbb{E}\left[H\right] &=\mathbb{E} \left[\mathbb{E}\left(H \, | \, x\right)\right] = \mathbb{E}\left[x^p\right] \\
				&= \int \frac{x^{p+k_1-1}}{2 \Gamma(k_1) \theta_1^{k_1}} e^{-\frac{x}{\theta_1}} \, \text{d}x
				+ \int \frac{x^{p+k_2-1}}{2 \Gamma(k_2) \theta_2^{k_2}} e^{-\frac{x}{\theta_2}} \, \text{d}x \\
				&= \frac{\Gamma(k_2) \Gamma(p+k_1) \theta_1^{p} + \Gamma(k_1) \Gamma(p+k_2) \theta_2^{p}}{2 \Gamma(k_1) \Gamma(k_2)}\\
				&= \frac{C}{2 \Gamma(k_1) \Gamma(k_2)},
			\end{align*}
	letting $C =  \Gamma(k_2) \Gamma(p+k_1) \theta_1^{p} + \Gamma(k_1) \Gamma(p+k_2) \theta_2^{p}$.	
	\end{enumerate}
	Combining the three components,
	\begin{align*}
		\alpha(\text{d}x) 
		&= \frac{2 \Gamma(k_1) \Gamma(k_2)}{C} \left( \frac{x^{p+k_1-1}}{2 \Gamma(k_1) \theta_1^{k_1}} e^{-\frac{x}{\theta_1}} 
				+ \frac{x^{p+k_2-1}}{2 \Gamma(k_2) \theta_2^{k_2}} e^{-\frac{x}{\theta_2}} \right) \, \text{d}x \\
		&= \underbrace{\frac{ \Gamma(k_2)\Gamma(p+k_1) \theta_1^{p+k_1}}{C \theta_1^{k_1}}}_{\varphi(p, k_{1:2}, \theta_{1:2})} 
		\underbrace{\frac{x^{p+k_1-1}}{\Gamma(p+k_1) \theta_1^{p+k_1}} e^{-\frac{x}{\theta_1}}}_{\text{Gamma}\left(p+k_1, \theta_1\right)} 
		+ \underbrace{\frac{\Gamma(k_1) \Gamma(p+k_2) \theta_2^{p+k_2}}{C \theta_2^{k_2}}}_{\varphi'(p, k_{1:2}, \theta_{1:2})} 
				\underbrace{\frac{x^{p+k_2-1}}{\Gamma(p+k_2) \theta_2^{p+k_2}} e^{-\frac{x}{\theta_2}}}_{\text{Gamma}\left(p+k_2, \theta_2\right)}  \, \text{d}x. 
	\end{align*}
	And now substituting back the expression $C$ in $\varphi$:
	\begin{align*}
		\varphi(p, k_{1:2}, \theta_{1:2}) &= \left(1 + \frac{\Gamma(k_1) \Gamma(p+k_2) \theta_2^{p}}{\Gamma(k_2) \Gamma(p+k_1) \theta_1^{p}} \right)^{-1}.
	\end{align*}
 Similarly, we can obtain $\varphi'(p, k_{1:2}, \theta_{1:2}) = 1 - \varphi(p, k_{1:2}, \theta_{1:2})$. Therefore, the anytime distribution $\alpha(\text{d}x)$ is the following mixture of two Gamma distributions:
	\begin{align*}
		\alpha(\text{d}x) &= \varphi(p, k_{1:2}, \theta_{1:2}) \, \text{Gamma}\left(k_1+p, \theta_1\right) 
		+ [1 - \varphi(p, k_{1:2}, \theta_{1:2})] \, \text{Gamma}\left(k_2+p, \theta_2\right).
	\end{align*}

\end{document}